\documentclass[lettersize,journal, compsoc]{IEEEtran}

\usepackage{graphicx}
\usepackage{amsmath}
\usepackage{cite}
\usepackage{url}
\usepackage{hyperref}
\hypersetup{
    colorlinks=true,
    linkcolor=blue,
    citecolor=blue,
    filecolor=magenta,      
    urlcolor=cyan
    }
    
\usepackage{graphicx}
\usepackage{array}
\usepackage{booktabs}
\usepackage{multirow}
\usepackage{graphicx}
\usepackage{subcaption}
\usepackage{cleveref}
\usepackage{listings}
\usepackage{xcolor}
\usepackage{float}

\begin{document}

%\title{Performance Evaluation of BBRv3 over Wi-Fi}

\title{TCP BBR Performance over Wi-Fi~6: AQM Impacts and Cross-Layer Insights}

\author{Shyam Kumar Shrestha, Shiva Raj Pokhrel and Jonathan Kua%
    \IEEEcompsocitemizethanks{
        \IEEEcompsocthanksitem S.K. Shrestha, S.R. Pokhrel and J. Kua are with the School of Information Technology, Deakin University, Geelong, Australia.
        Email: \{shyam.shrestha, shiva.pokhrel, jonathan.kua\}@deakin.edu.au}}

\IEEEtitleabstractindextext{
\begin{abstract} 

We evaluate TCP~BBRv3 on Wi-Fi~6 home networks under modern AQM schemes using a fully wireless testbed and a simple cross-layer model linking Wi-Fi scheduling, router queuing, and BBRv3’s pacing dynamics. Comparing BBR Internet traffic with CUBIC across different AQMs (FIFO, FQ-CoDel, and CAKE) for uplink, downlink, and bidirectional traffic, we find that FIFO destabilizes pacing and raises delay, often letting CUBIC dominate; FQ-CoDel restores fairness and controls latency; and CAKE delivers the best overall performance by keeping delay low and aligning BBRv3’s sending and delivered rates. We also identify a Wi-Fi-specific effect where CAKE’s rapid queue draining, while improving pacing alignment, can trigger brief retransmission bursts during BBRv3’s bandwidth probes. These results follow from the interaction of variable Wi-Fi service rates, AQM delay control, and BBRv3’s inflight limits, leading to the practical guidance to use FQ-CoDel or CAKE and avoid unmanaged FIFO in home Wi-Fi, with potential for Wi-Fi-aware tuning of BBRv3’s probing.

\end{abstract}

}

\markboth{IEEE TRANSACTIONS ON MOBILE COMPUTING}%
{Shell \MakeLowercase{\textit{et al.}}: %A Sample Article Using IEEEtran.cls for IEEE Journals
}
\IEEEpubid{0000--0000/00\$00.00~\copyright~2025 IEEE}

\maketitle

\section{Introduction} \label{introduction}

Wi-Fi has become the dominant access technology for residential and small-office connectivity, with household penetration exceeding 85\% in developed regions. This growth is amplified by the rapid expansion of Internet of Things (IoT) ecosystems: the average home now operates 22.6 connected devices, a number projected to increase by 38\% by 2027. Consequently, residential networks have evolved into dense, heterogeneous, and highly interactive wireless environments. Multiple uplink, downlink, and bidirectional flows compete for airtime while experiencing contention, interference, and time-varying channel conditions that significantly influence transport-layer performance~\cite{pokhrel2018modeling,shrestha2025visualizing}. Under these conditions, the Transmission Control Protocol (TCP) congestion control algorithm (CCA) plays a central role in determining overall system throughput, latency, and fairness among competing traffic flows~\cite{8365841}.
\begin{figure}[t]
\centering
\includegraphics[width=0.5\linewidth]{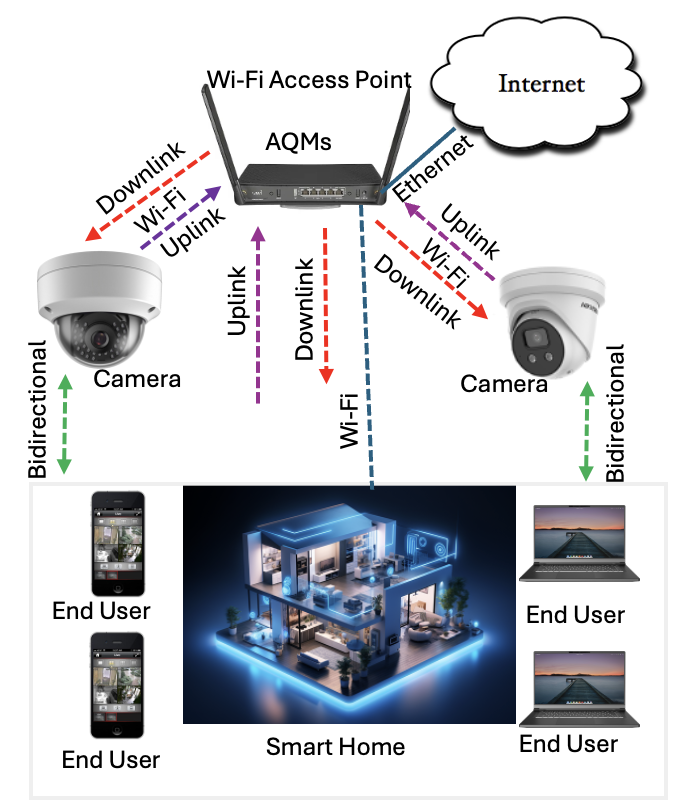}
\caption{A residential Wi-Fi network with multiple IP cameras uploading surveillance footage and end-user devices consuming downlink content. Uplink, downlink, and bidirectional traffic compete for airtime at the access point (AP), creating realistic congestion conditions for evaluating BBRv3 and CUBIC under PFIFO, FQ-CoDel and CAKE.}
\label{fig:usecase_fig}
\end{figure}
Designing a universally robust congestion-control mechanism for Wi-Fi remains challenging due to the medium’s intrinsic variability: random losses, fluctuating physical-layer rates, dynamic Multiple-User Orthogonal Frequency-Division Multiple Access (MU-OFDMA) scheduling, and shallow buffer constraints~\cite{du2024revisiting}. Loss-based CCAs such as CUBIC (default in Linux) interpret packet loss as a congestion signal~\cite{ha2008cubic}. However, in Wi-Fi, losses frequently arise from non-congestion phenomena such as collisions or channel fading, causing CUBIC to reduce its sending rate unnecessarily and resulting in suboptimal throughput and fairness~\cite{shrestha2024fairness}.

To address these limitations, Google introduced the model-based Bottleneck Bandwidth and Round-trip propagation time (BBR) algorithm~\cite{cardwell2016bbr} in 2016, followed by version 2 (BBRv2~\cite{cardwell2018bbr}) in 2019, and version 3 (BBRv3~\cite{cardwell2023bbrv3}) in 2023. BBRv3 aims to achieve more consistent and fair performance by explicitly estimating bottleneck bandwidth and propagation delay, pacing traffic accordingly. Prior studies demonstrate that BBRv3 improves coexistence, responsiveness, and fairness in wired and explicit congestion notification (ECN)-enabled environments~\cite{zeynali2024bbrv3,shrestha2025visualizing,gomez2024evaluating}. However, these evaluations primarily focus on wired networks; the behavior of BBRv3 in fully wireless, contention-prone Wi-Fi environments remains largely unexplored~\cite{shrestha2025visualizing}. In such environments, stochastic medium access, random collisions, and queue buildup interact in complex ways that existing analytical models do not fully capture.

Modern Active Queue Management (AQM) mechanisms particularly Flow-Queue Controlled Delay (FQ-CoDel)~\cite{hoeiland2018flow} and Common Applications Kept Enhanced (CAKE)~\cite{hoiland2018piece} are increasingly deployed in home gateways to mitigate bufferbloat and improve fairness. Recent studies have shown that AQM configurations can significantly influence the behavior of BBR family algorithms~\cite{cardwell2017bbr,cardwell2018bbr,cardwell2023bbrv3, shrestha2025visualizing}. Yet, to date, no systematic evaluation has examined BBRv3 under packet first-in first-out (PFIFO), FQ-CoDel, and CAKE in a real-world Wi-Fi testbed, nor provided analytical insights into how queuing disciplines shape pacing, delay, retransmissions, and coexistence with CUBIC.

A practical use-case illustrating these challenges is a residential smart-home surveillance scenario (Fig.~\ref{fig:usecase_fig}), where multiple Internet Protocol (IP) cameras continuously upload high-bitrate video while users simultaneously consume downlink content. These uplink, downlink, and bidirectional flows share the same Wi-Fi bottleneck, stressing both MAC-layer scheduling and transport-layer congestion control. Understanding the behaviors of modern CCAs in such contention-heavy environments is critical for achieving fairness, stability, and predictable performance.

%\subsection*{Contributions}
To address these gaps, this paper presents the first systematic experimental evaluation of TCP BBRv3 over a fully wireless IEEE~802.11ax (Wi-Fi~6)  testbed under three AQM disciplines (PFIFO, FQ-CoDel, and CAKE) and across uplink, downlink, and bidirectional traffic flows. The main contributions are as follows:

\begin{itemize}
    \item \textbf{Cross-layer modeling and analysis:}  
    We develop a unified analytical framework that couples MU-OFDMA MAC scheduling, AQM characteristics, and BBRv3 fluid dynamics. This framework interprets and explains experimental observations such as throughput oscillations and delay patterns.  

    \item \textbf{Comprehensive real-world evaluation:}  
    We perform systematic measurements across uplink, downlink, and bidirectional traffic. This provides novel insights into pacing delivery interactions, retransmissions, jitter, and fairness, highlighting how different AQMs (PFIFO, FQ-CoDel and CAKE) affect BBRv3 and CUBIC in real Wi-Fi~6 environments.  

    \item \textbf{Experimental testbed design and implementation:}  
    We design and deploy a flexible Wi-Fi~6 testbed using a Mikrotik router, enabling configurable queuing mechanisms. This setup supports reproducible experimentation and provides critical insights into the impact of queue management on pacing, retransmissions, and flow fairness.
\end{itemize}

The remainder of the paper is organized as follows. Section \ref{sec:background} provides background of the TCP BBR including BBRv3, AQMs, IEEE 802.11 MU-OFDMA and, reviews related work on TCP congestion control in Wi-Fi and AQM systems. Section \ref{sec:system_model} presents the modeling of MU-OFDMA throughput, BBR fluid model and AQMs. Section~\ref{experimentaltestbed} describes the experimental testbed, traffic scenarios and measurement methodology. Section~\ref{analysis} presents the evaluation results, including retransmission performance, latency, and pace. Section~\ref{conclusion} concludes with recommendations and future directions.

\section{Background and Related Work} \label{sec:background}

The rapid proliferation of connected devices and the increasing dependency on Wi-Fi make congestion control a critical component of modern wireless networks~\cite{du2024revisiting,shrestha2024fairness}. In particular, Wi-Fi~6 introduces OFDMA-based uplink multi-user (MU) scheduling that significantly reshapes MAC-layer dynamics, requiring transport-layer algorithms to adapt to variable service rates and contention patterns. This section reviews the foundational concepts relevant to this study: the evolution of BBR congestion-control algorithm (focusing on BBRv3), the role of AQ schemes in wireless performance, and existing analytical models for MU--OFDMA networks.

\subsection{BBRv3}
\label{bbrv3}

BBRv3 is the most recent evolution of TCP BBR CCAs. It extends the model-based framework introduced by BBRv1~\cite{cardwell2017bbr} and refined in BBRv2~\cite{cardwell2018bbr} by improving robustness, fairness, and coexistence with loss-based congestion control. The algorithm is specified in the latest IETF draft~\cite{cardwell2023bbrv3}, which provides detailed updates to pacing behavior, inflight limits, and congestion-response logic.

BBR operates by independently estimating two key properties of a network path: the bottleneck bandwidth ($BtlBw$) and the minimum round-trip propagation time ($RTprop$). BBRv1 used fixed $pacing\_gain$ cycles to probe for bandwidth, but its aggressive gain values and lack of explicit loss response often leads to persistent queuing and unfairness\cite{zeynali2024bbrv3,gomez2024evaluating,shrestha2025visualizing}. BBRv2 addressed several of these issues by introducing \emph{inflight\_hi} and \emph{inflight\_lo} limits, loss/ECN-based adjustments, and a structured \textit{ProbeBW} state machine~\cite{cardwell2018bbr,zeynali2024bbrv3,shrestha2025visualizing}. 

\begin{figure} [h]
    \centering
    \includegraphics[scale=0.1]{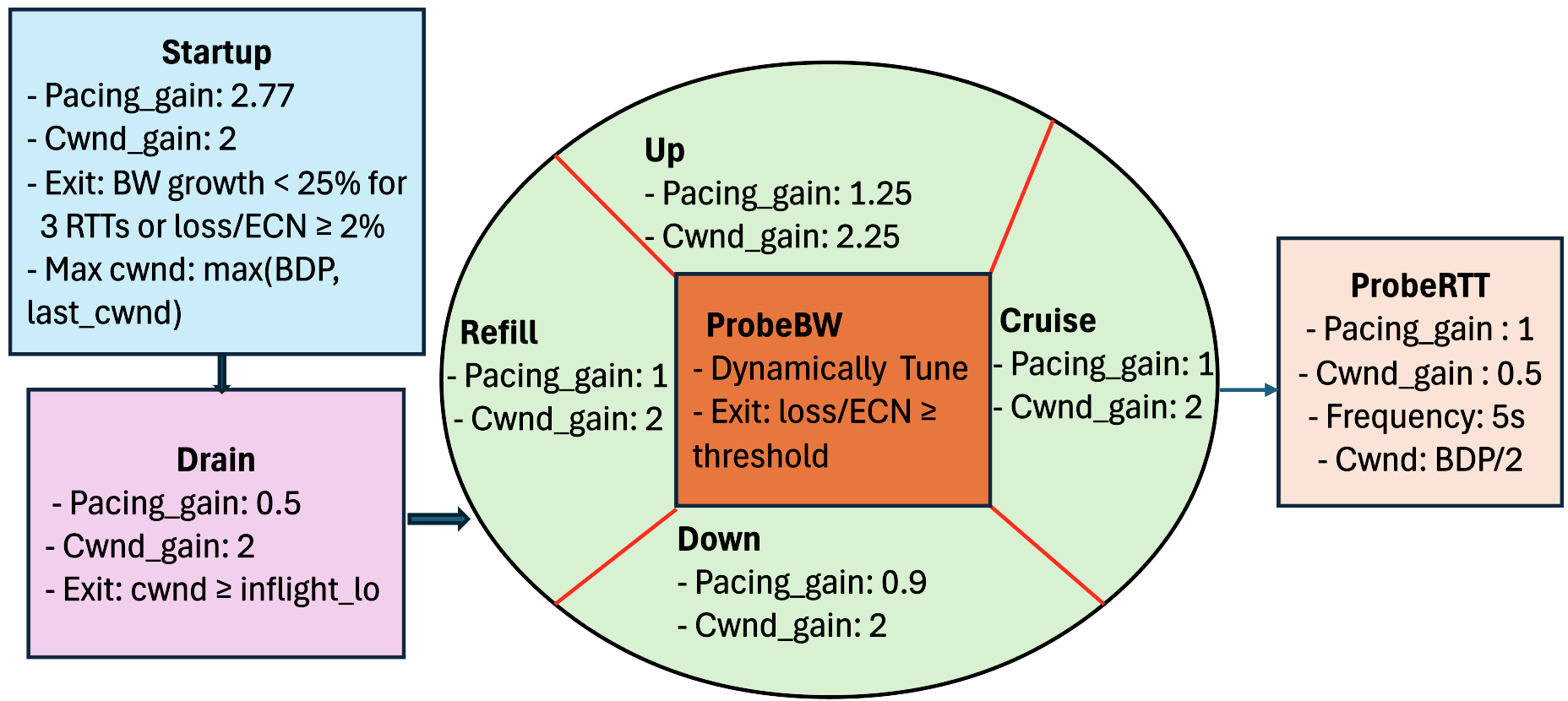}
    \caption{BBRv3 state diagram. After \textit{Startup} and \textit{Drain}, the sender repeatedly
cycles through the \textit{ProbeBW} sub-phases \textit{(Up, Down, Refill, Cruise)}, adjusting
$pacing\_gain$ and $cwnd\_gains$ to track bottleneck bandwidth. \textit{ProbeRTT} is entered 
periodically ($\approx 5\,$s) to refresh the minimum RTT using a $inflight\_lo$ probing window.}

    \label{fig:bbr_states}
\end{figure}

% \subsection{Wireless Testbed Architecture}
% \label{sec:testbed}

\begin{table*}[!t]
\centering
\caption{Comparison of BBR Versions: Key Parameters and Roles in Wi-Fi Context.}
\label{tab:bbr_fullwidth_wifi}
\scriptsize
\renewcommand{\arraystretch}{1.4}
\begin{tabular}{|p{2cm}|p{2cm}|p{2.5 cm}|p{2.7 cm}|p{6 cm}|}
\hline
\textbf{Parameter}~\cite{cardwell2017bbr,cardwell2018bbr,cardwell2023bbrv3} & 
\textbf{BBRv1}~\cite{cardwell2017bbr} & 
\textbf{BBRv2}~\cite{cardwell2018bbr} & 
\textbf{BBRv3}~\cite{cardwell2023bbrv3} & 
\textbf{Purpose / Role in Wi-Fi context}\\
\hline
\multicolumn{5}{|c|}{\textbf\textit{{Startup}}}\\
\hline
$pacing\_gain$ & 2.89 & 2.89 & 2.00–2.77 & Rapid bandwidth probing avoiding queue buildup \\
\hline
$cwnd\_gain $& 2.89 & 2.89 & 2.00 & Control inflight growth to avoid Wi-Fi bufferbloat\\
\hline
Exit condition & BW growth$<25\%$ for 3 RTTs & BW growth$<25\%$ for 3 RTTs or loss/ECN $\geq 2\%$ & BW growth$<25\%$ for 3 RTTs or loss/ECN $\geq 2\%$
& Exit when bandwidth plateaus or loss appears, even if loss is from interference \\
\hline
Max $cwnd$ & -- & Prev. observed max & $max(BDP, last\_cwnd)$ & Limit growth to prevent bufferbloat in AP queues  \\
\hline
\multicolumn{5}{|c|}{\textbf\textit{{Drain}}} \\
\hline
$pacing\_gain$ & 0.75 & 0.75 & 0.50 & \textit{Drain}  excess queue to reduce inflated RTT from shared-medium delays \\
\hline
Exit condition &  $cwnd$ $\geq$ $1.BDP$ &  $cwnd$ $\geq$ $inflight\_lo$ & $cwnd $$\geq$ $inflight\_lo$  & Switch to steady state once inflight matches fluctuating Wi-Fi capacity \\
\hline
\multicolumn{5}{|c|}{\textbf\textit{{ProbeBW}}} \\
\hline
Sub-phases & N/A & \textit{Cruise, Refill, Up, Down} & \textit{Cruise, Refill, Up, Down} (dynamically tuned) & Probe adaptively to track bandwidth spikes/drops from contention \\
\hline
$pacing\_gain $& 1.25,0.75,1... & $Up$: 2.0, $Down$: 0.75 & $Up$ : 1.25, $Down:$ 0.90 & Probe without triggering excessive retries or collisions \\
\hline
$cwnd\_gain $& Follows $pacing\_gain $& $Up$ = 2.0, others phase-based & $Up$ = 2.25, tuned & Balance rate and window to avoid overfilling small Wi-Fi buffers \\
\hline
Exit condition & $cwnd$ $\geq$ $pacing\_gain$ $\times$BDP or loss & loss/ECN $\geq$ threshold & loss/ECN $\geq$ threshold & Leave probing when persistent loss/ECN indicates congestion \\
\hline
\multicolumn{5}{|c|}{\textbf\textit{{ProbeRTT}}} \\
\hline
Frequency & 10 s & 5 s & 5 s & Periodically measure delay to separate congestion from contention \\
\hline
$cwnd$ & 4 pkts & $BDP/2$ & $BDP/2$ & Lower inflight for accurate RTT sampling despite queueing \\
\hline
Duration & ~200 ms or 1 RTT & 1 RTT or 200 ms & 1 RTT or 200 ms & Keep short to avoid throughput drop during access delays \\
\hline
\multicolumn{5}{|c|}{\textbf{Congestion Limits}} \\
\hline
$inflight\_hi$ & Not defined & For fairness control & Tuned dynamic & Prevent overshooting AP buffer \\
\hline
$inflight\_lo$ & Not used & \textit{ProbeBW:Down} & \textit{ProbeBW:Down} & Maintain progress despite random interference losses \\
\hline
Random probe & Fixed 8-cycle & Adaptive 2–3 s & Adaptive tuned & Randomize probing to prevent burst synchronisation on shared channels \\
\hline
\end{tabular}
\end{table*}

BBRv3 retains the core architecture of BBRv2 but introduces important refinements~\cite{cardwell2023bbrv3,cardwell2024bbr}. First, it adopts tuned $pacing\_gain$ values that reduce the aggressiveness of both \textit{Startup} and \textit{ProbeBW}. In contrast to the fixed $\approx 2.89$ gain used in BBRv1 and BBRv2, BBRv3 applies a more conservative range of approximately $2.0$--$2.77$ during \textit{Startup}, and employs updated \textit{Up} and \textit{Down} gains in \textit{ProbeBW} to better regulate bandwidth probing and queue draining. Second, BBRv3 strengthens congestion-window $(cwnd)$ control by retaining the BBRv2 constraint $cwnd \le \max({BDP}, {last\_cwnd})$ while refining $cwnd $ growth in response to recently observed loss. In particular, loss events update ${inflight\_hi}$, enabling the algorithm to cap inflight data when persistent congestion is detected. Third, BBRv3 preserves the structured four-phase \textit{ProbeBW} state machine ($Cruise$, $Refill$, $Up$, $Down$) introduced in BBRv2 but adjusts $pacing\_gain$ and transition logic as shown in Figure~\ref{fig:bbr_states} . Loss or ECN signals may terminate \textit{ProbeBW} early, improving coexistence with loss-based congestion control. Fourth, similar to BBRv2, BBRv3 periodically enters \textit{ProbeRTT} to refresh its estimate of $RTprop$, temporarily reducing inflight to approximately ${BDP}/2$ to facilitate accurate RTT sampling. Finally, BBRv3 employs a more controlled probing cadence, reducing the likelihood of flow synchronization and mitigating burst-induced queuing.

Table~\ref{tab:bbr_fullwidth_wifi} summarizes key parameter differences among BBRv1, BBRv2 and BBRv3, based on the published specifications~\cite{cardwell2017bbr,cardwell2018bbr,cardwell2023bbrv3}. The progressive adjustments across versions illustrate the algorithm’s shift from fixed and aggressive probing in BBRv1 to the more conservative, loss-aware, and fairness-oriented behavior in BBRv3.

Overall, BBRv3 preserves the core BBR principle of operating near the estimated bottleneck BDP while addressing several practical shortcomings of earlier versions, particularly regarding fairness, congestion response, and robustness in diverse network environments.

\subsection{MU--OFDMA and IEEE~802.11ax MAC Behaviour}
IEEE~802.11ax introduces MU-OFDMA, enabling simultaneous transmissions from multiple stations using orthogonal resource units (RUs). This alters collision patterns, backoff behavior, and aggregation opportunities, directly affecting transport-layer performance. The complex interactions between the new scheduled access and legacy contention-based access must be carefully considered, particularly for deriving accurate MAC-layer service rates \cite{behara2022performance, magrin2023performance}.

The Trigger Frame (TF) cycle, consisting of Trigger Frame Response (TF-R), Buffer Status Report (BSR), Trigger Frame (TF), Physical Protocol Data Unit (PPDU), Multi-Station Block Acknowledgment (MS-BACK), and the associated Short Interframe Space (SIFS) intervals, defines the fundamental time unit for MU scheduling~\cite{behara2022performance, magrin2023performance,saldana2017frame } . Within this structure, throughput depends on the number of aggregated packets, backoff behavior, and collision probability requiring accurate analytical modeling to derive per-STA service rates. Fixed-point models based on Bianchi’s formulation~\cite{bianchi2002performance} provide tractable approximations of per-STA attempt probabilities and MAC-layer service rates.

\subsection{Active Queue Management in Wi-Fi Networks}
\label{sec:aqm_background}

AQM is essential for controlling delay, congestion signalling and fairness in Wi-Fi networks, particularly in dense home networking environments. While CCAs such as CUBIC and BBRv3 regulate sender behaviors, their performance depends strongly on the queue discipline implemented at the Wi-Fi access point (AP). Modern AQMs mitigate the persistent buffer build-up commonly observed with un-managed FIFO queues~\cite{shrestha2025visualizing,hoeiland2018flow,hoiland2018piece}.

Conventional FIFO (drop-tail) queues admit packets until the buffer is full, after which all excess packets are dropped. This behavior frequently leads to long queues, inflated latency and delayed congestion feedback under contention~\cite{misra2000fluid,bianchi2002performance,floyd2008internet,scherrer2022model}. Thus, FIFO provides a useful baseline but performs poorly in scenarios with multiple active stations or high offered load.

FQ-CoDel addresses these limitations by combining per-flow queuing with the CoDel AQM algorithm~\cite{hoeiland2018flow,nichols2018controlled}. Each flow is isolated in its own queue and scheduled in a fair manner, while CoDel regulates sojourn time to maintain low delay. Empirical studies show that FQ-CoDel achieves substantially better latency and fairness than FIFO in home-gateway and residential Wi-Fi settings~\cite{hoiland2018piece}.

CAKE extends FQ-CoDel with features tailored for residential and wireless links. It incorporates per-host fairness, bandwidth shaping, and improved overhead compensation, alongside adaptive marking based on queue occupancy and delay~\cite{hoiland2018piece,nichols2012controlling,pan2013pie}. Recent work demonstrates that CAKE delivers favorable latency-throughput trade-offs for TCP traffic in Wi-Finetworks~\cite{shrestha2025visualizing}.

Since AQM mechanisms determine how congestion signals are generated, they play a direct role in shaping the behavior of modern CCAs. For delay-sensitive algorithms such as BBRv3, the choice of queue discipline significantly affects bandwidth estimation, RTT measurements, and fairness outcomes in multi-device Wi-Fi deployments~\cite{shrestha2025visualizing,domanski2021iot}. Hence, understanding AQM behavior is central to cross-layer design and analysis of TCP BBR performance over Wi-Fi~6.

\subsection{Related Work and Research Gaps}
\label{related}

Recent work on congestion control, queue management, and IEEE~802.11ax MAC modeling forms the foundation for understanding modern transport performance in wireless networks. While each of these components has been studied extensively in isolation, their interaction particularly in the context of BBRv3 operating over AQM-managed Wi-Fi 6 remains under-explored. This section reviews the state-of-the-art across these domains and identifies the gaps that motivate our study.

\subsubsection{TCP Congestion Control and BBR Evolution}

BBRv1 introduced a fundamentally new, model-based approach to congestion control by explicitly estimating the \textit{BtlBw} and \textit{RTprop} to pace at the bottleneck rate~\cite{cardwell2017bbr}. Although highly influential, subsequent studies reported persistent queues, latency inflation, and coexistence unfairness when BBRv1 competed against loss-based CCAs such as CUBIC~\cite{zeynali2024bbrv3,gomez2024evaluating,shrestha2025visualizing,shrestha2024fairness}.

BBRv2 addressed several of these issues through ECN- and loss-driven window reductions and a revised \textit{ProbeBW} cycle~\cite{cardwell2018bbr}. BBRv3 further refines $pacing\_gain$, \textit{startup} gain ranges, and in-flight bounding to improve fairness and to reduce persistent queue buildup~\cite{cardwell2023bbrv3}. Parallel to these protocol refinements, fluid-flow analyses have provided formal models of BBR’s rate and $cwnd$ dynamics, establishing a rigorous analytical basis for understanding its state transitions and steady-state behaviors~\cite{scherrer2022model,inoue2024fluid}.

Despite these advances, empirical evaluations of BBRv3 remain largely confined to wired, datacenter or high-speed wide-area environments, focusing on metrics such as fairness under mixed CCAs, ECN responsiveness, and behavior at 1-10\,Gbps line rates~\cite{bless2025insights,yang2023optimization,gomez2024evaluating}. To date, no study has systematically evaluated BBRv3 on real IEEE~802.11ax Wi-Fi systems, where PHY-rate variability, collision-induced losses, and frame aggregation create congestion signals that are very different from those present in wired paths.

\subsubsection{Active Queue Management Systems}

Queue management plays a central role in shaping transport-layer behaviors. Conventional drop-tail queues accept packets until the buffer saturates, resulting in bufferbloat and excessive queuing delay, which is especially problematic in Wi-Fi due to contention and feedback latency~\cite{floyd2008internet}. 

Modern AQMs such as CoDel~\cite{nichols2012controlling} and FQ-CoDel~\cite{pokhrel2018modeling} mitigate bufferbloat through delay-based dropping and flow isolation. CAKE extends this paradigm by adding per-host fairness, bandwidth shaping, and Wi-Fi-aware overhead compensation~\cite{hoiland2018piece}. These mechanisms are now deployed on home routers and are widely used in practice.

Analytical models for AQM behaviors particularly fluid models describing queue occupancy, sojourn time, and marking/dropping probability have been well-established since the proposal of the Random Early Detection (RED) model~\cite{misra2000fluid}. Empirical studies further confirm that FQ-CoDel and CAKE significantly reduce latency in wireless networks~\cite{domanski2021iot, shrestha2025visualizing,9525028}. However, no prior work has examined how these AQMs interact with BBRv3’s pacing, $cwnd$ bounding, and \textit{ProbeBW} behavior under MU-OFDMA Wi-Fi scheduling. This gap is particularly important because modern Wi-Fi APs increasingly rely on AQM to control latency under load.

\subsubsection{MU-OFDMA MAC Modeling and Cross-Layer Performance}

Transport performance over 802.11 networks is inherently a cross-layer challenge, shaped by how MAC-layer service opportunities influence transport-layer queuing and congestion signals. The analytical foundation for wireless MAC modeling is the Bianchi Markov chain~\cite{bianchi2002performance}, which characterizes collision probability, attempt probability, and throughput under contention.

With the introduction of IEEE~802.11ax, several works have extended this model to incorporate MU-OFDMA TF cycles, enabling per-STA service-rate derivation~\cite{pokhrel2018modeling,behara2022performance,magrin2023performance}. These models quantify the timing structure and resource-unit (RU) allocation of the UL OFDMA cycle, providing a basis for analyzing service variability insights that are directly used in our MU-OFDMA model. Following this approach, we model MU-OFDMA throughput in our work based on the framework proposed in~\cite{behara2022performance}, allowing us to capture per-STA service rates and queue dynamics under realistic traffic conditions. Recent physical-layer studies have additionally examined RU-level packet loss characteristics through correlation-based and multi-dimensional Markov-chain models~\cite{zhang2025packet}. These works reveal that frequency-selective fading induces non-trivial temporal and frequency-domain correlations in RU reliability. However, they focus on PHY-layer packet-loss structure and do not analyze its interaction with queuing, AQM dynamics, or congestion control algorithms such as BBRv3.

Recent studies also highlight the practical advantages and limitations of MU-OFDMA in IEEE~802.11ax. While MU-OFDMA reduces channel access overheads under low traffic conditions, it can incur extra overheads under saturated traffic, potentially reducing throughput compared to single-user (SU) transmissions~\cite{lee2025enriching}. Despite this limitation, MU-OFDMA can enhance overall Wi-Fi network capacity by flexibly allocating small RUs over frequency selective channels.
% Measurement studies reveal that commercial APs currently underutilize RU allocation due to insufficient consideration of detailed channel characteristics. To address this, ChORUS proposes a standard-compliant framework for optimal RU allocation that exploits frequency diversity across users. Trace-driven simulations and proof-of-concept experiments show that ChORUS significantly improves throughput performance over SU transmission, demonstrating MU-OFDMA's capacity benefits in real-world deployments.

Wireless evaluations of BBR remain limited and primarily focus on BBRv1, BBR-P and BBR-n over older 802.11n/ac systems~\cite{grazia2020bbrp,miyazawa2020performance,ahsan2023tcp}. To the best of our knowledge, there is no study that investigate the performance of BBRv3 in Wi-Fi~6, and how its model-based pacing interacts with the dynamic service rates and queue behaviors introduced by MU-OFDMA and AQM.

\subsubsection{Research Gaps}

While prior research provides strong foundations across BBR design, AQM behaviors, and MU-OFDMA MAC modeling, there remains several critical research gaps as follows:

\begin{itemize}
    \item \textbf{Empirical Gap: BBRv3 on real Wi-Fi 6 with modern AQMs.}  
    No systematic evaluation of BBRv3 exists across PFIFO, FQ-CoDel, and CAKE on real Wi-Fi 6 hardware, despite widespread deployment of these AQMs in home routers.

    \item \textbf{Performance-Anomaly Gap: Interaction between CAKE and BBRv3.}  
    Prior wired analyses do not observe the retransmission anomaly we identify, where CAKE’s rapid queue draining interacts with BBRv3’s \textit{ProbeBW} dynamics, creating pacing delivery misalignment.

    \item \textbf{Analytical Gap: Lack of a unified cross-layer model.}  
    Existing models independently capture MAC service rates, AQM queue dynamics, or BBR behaviors, but none integrate all three aspects in analyzing how MU-OFDMA variability shapes BBRv3’s in-flight and pacing evolution.
\end{itemize}

The table in~\autoref{tab:related_works} summarizes recent literature in the field and highlights how our work differs from prior studies.

\begin{table*}[!t]
\centering
\caption{Summary of Recent Literature on AQM, BBR, and IEEE~802.11ax Cross-Layer Performance}
\label{tab:related_works}
\begin{tabular}{|p{4.2cm}|p{2.3cm}|p{1.6cm}|p{4.40cm}|p{2.9cm}|}
\hline
\textbf{Study} & \textbf{AQM Considered} & \textbf{BBR Variant} & \textbf{Cross-Layer Modeling} & \textbf{IEEE~802.11ax Support} \\
\hline

\cite{zeynali2024bbrv3,gomez2024evaluating,bless2025insights} 
& No & Yes & No & Wired / Simulation \\
\hline

\cite{shrestha2025visualizing}
& Yes & Yes & No & Yes \\
\hline

\cite{grazia2020bbrp,miyazawa2020performance,du2024revisiting} 
& No & Yes & No & Earlier IEEE~802.11ac \\
\hline

\cite{scherrer2022model, inoue2024fluid}
& No & Yes & BBR Fluid Model & No \\
\hline

\cite{floyd2008internet,hoiland2018piece,9525028,nichols2018controlled,nichols2012controlling,domanski2021iot}
& Yes & No & AQM Modeling & No \\
\hline

\cite{pokhrel2018modeling,behara2022performance,magrin2023performance,bianchi2002performance,lee2025enriching,zhang2025packet,saldana2017frame}
& No & No & MU--OFDMA / OFDMA  Modeling & Yes \\
\hline

\textbf{Our Work}
& Yes & Yes & Yes & Yes \\
\hline

\end{tabular}
\end{table*}

% \noindent
% \textbf{Contributions.}  
% This study addresses these gaps through:
% \begin{itemize}
%     \item the first systematic experimental evaluation of BBRv3 and CUBIC across PFIFO, FQ-CoDel, and CAKE in real Wi-Fi~6 environments;
%     \item the discovery and detailed characterization of a novel CAKE--BBRv3 retransmission anomaly; and
%     \item a unified analytical framework that couples MU-OFDMA MAC scheduling, AQM behavior, and BBRv3 fluid dynamics to explain observed performance interactions.
% \end{itemize}

\section{System Model}
\label{sec:system_model}
 
In this section, we present the cross-layer system model used in this work. All symbols introduced throughout this section are summarized in the notation Table \ref{tab:symbols_all_global}. We first model the Wi-Fi 6 MU-OFDMA MAC and derive per-STA service rates. Building on the MAC model, we present a fluid formulation of BBRv3 congestion-control dynamics, followed by the end-to-end RTT and queue dynamics that couple transport and MAC layers. Finally, we describe our AQM models (PFIFO, FQ-CoDel, CAKE) and how they integrate into the cross-layer framework.

\begin{table*}[!t]
\centering
\footnotesize % Reduce font size
\setlength{\tabcolsep}{2pt}
\caption{Global notations and symbols used throughout the MU--OFDMA, BBRv3, cross-layer, and AQM modeling sections. Symbols are listed alphabetically across six columns for reference.}
\label{tab:symbols_all_global}
\renewcommand{\arraystretch}{1.10}
\setlength{\tabcolsep}{2pt} % makes columns narrower
\begin{tabular}{ll ll ll}

\toprule
\textbf{Symbol} & \textbf{Description} &
\textbf{Symbol} & \textbf{Description} &
\textbf{Symbol} & \textbf{Description} \\
\midrule

$B_k$ & Avg.\ backoff slots at stage $k$ &
$\Delta(p_{\pi_i})$ & Loss-response intensity &
$d_{\mathrm{bsr}}$ & Buffer Status Report duration \\

$d_{\mathrm{mb}}$ & Multi-STA Block ACK duration &
$d_{\mathrm{ppdu}}$ & PPDU duration &
$d_{\mathrm{tf}}$ & Trigger-frame duration \\

$d_{\mathrm{tfr}}$ & TF-R frame duration &
$d_{\mathrm{sifs}}$ & SIFS duration &
$d_i^{\mathrm{fifo}}(t)$ & FIFO per-flow drop rate \\

$E[X]$ & Expected slot duration &
$f_i$ & Per-attempt failure probability &
$fbo$ & Fixed backoff slots \\

$F_{\mathrm{cake}}$ & CAKE AQM marking function &
$g_{\mathrm{hi}}$ & High-gain probe factor &
$h(i)$ & Host ID of flow $i$ \\

$m_i^{\mathrm{cake}}(t)$ & CAKE drop/mark probability &
$m_i^{\mathrm{crs}}$ & Cruise indicator (0, 1) &
$mbo$ & Max backoff stage \\

$n_j$ & Number of STAs in class $j$ &
$n_{\mathrm{ra}}$ &  Random-access STAs &
$n_p$ & Packets available for aggregation \\

$P_0$ & Idle-slot probability &
$P_{\mathrm{Agg}}^{\mathrm{Max}}$ & Hardware aggregation limit &
$P_{\mathrm{Agg}}^{\mathrm{Trf}}$ & Max aggregated packets per MU-TF \\

$P_f$ & Failed-slot probability &
$P_s$ & Successful-slot probability &
$p_{\mathrm{agg}}$ & Aggregated packets per MU \\

$p_i(t)$ & AQM/MAC drop probability &
$p_{\mathrm{drop}}$ & Drop/mark function &
$p_{\mathrm{phy}}$ & PHY-layer packet error probability \\

$p_{\mathrm{th}}$ & Loss threshold &
$p_{\pi_i}$ & Packet-loss probability &
$q_i(t)$ & Queue occupancy \\

$q_{\mathrm{tot}}(t)$ & Total queue occupancy &
$r_A$ & Random-access resource factor &
$s_A$ & Scheduled MU transmissions per TF \\

$s_h$ & Header bits &
$s_i^{\mathrm{cake}}(t)$ & CAKE service allocation &
$s_i^{\mathrm{fifo}}(t)$ & FIFO service allocation \\

$s_i^{\mathrm{fq}}(t)$ & FQ-CoDel per-flow service &
$s_p$ & Payload size (bits) &
$s_t$ & Trailer bits \\

$T_c$ & Collision-slot duration &
$T_{\mathrm{Phy}}^{\mathrm{MU}}$ & MU PHY overhead &
$T_r$ & STA PHY rate \\

$T_s$ & Successful-slot duration &
$t_d$ & TXOP duration &
$t_i^{\mathrm{pbw}}$ & RTT sample during \textit{ProbeBW}  \\

$t_s$ & TF-cycle duration &
$\tau_i(t)$ & Observed RTT &
$\tau_i^{\min}$ & Minimum RTT \\

$\Theta$ & Aggregate MU-OFDMA throughput &
$\Theta_i$ & Per-STA service rate &
$\Theta_i(t)$ & Throughput under drops/marks \\

$v_i$ & Instantaneous inflight data &
$w_i^{\mathrm{hi}}$ & High-bound congestion window &
$w_i^{\mathrm{lo}}$ & Low-bound congestion window \\

$w_i^{\mathrm{pbw}}$ & \textit{ProbeBW} window &
$w_i^{\mathrm{prt}}$ & \textit{ProbeRTT} window &
$\overline w_i$ & Base congestion window \\

$\dot{w}_i^{\mathrm{lo}}$ & Derivative of low-bound window &
$\alpha_i^{\mathrm{cake}}(t)$ & CAKE service share \\

$\alpha_i^{\mathrm{fq}}(t)$ &  FQ-CoDel weight &
$\omega_i(t)$ & CAKE dynamic weight &
$\phi_{\mathrm{host}}(h(i))$ & Per-host fairness factor \\

$\psi_i$ & Per-flow fairness weight &
$\sigma$ & Idle-slot duration &
$\varsigma(\cdot)$ & Smooth activation/transition function \\

$x_i(t)$ & Active sending rate &
$x_i^{\mathrm{pbw}}(t)$ & Sending rate during \textit{ProbeBW}  &
$x_i^{\mathrm{prt}}(t)$ & Sending rate during \textit{ProbeRTT} \\
$\mathrm{DIFS},\mathrm{SIFS}$ & MAC interframe spacing & & & & \\

\bottomrule
\end{tabular}
\end{table*}

\subsection{MU-OFDMA Throughput Modelling}
\label{sec:muofdma}

To model MU-OFDMA throughput under IEEE~802.11ax we adopt the trigger-frame (TF) cycle as the natural time unit \cite{behara2022performance}. The TF-cycle duration is:
\begin{equation}
\label{eq:ts_cycle}
t_s = d_{\mathrm{tfr}} + d_{\mathrm{bsr}} + d_{\mathrm{tf}} + d_{\mathrm{ppdu}} + d_{\mathrm{mb}} + 4 d_{\mathrm{sifs}},
\end{equation}
where $d_{\mathrm{tfr}}$, $d_{\mathrm{bsr}}$, $d_{\mathrm{tf}}$, $d_{\mathrm{ppdu}}$, and $d_{\mathrm{mb}}$ are the durations of TF-R, BSR, TF, PPDU, and MS-BACK frames, respectively, and $d_{\mathrm{sifs}}$ is the SIFS duration. Equation~\eqref{eq:ts_cycle} establishes the time base used throughout the analysis.

The aggregate MU-OFDMA throughput across all STAs is written as:
\begin{equation}
\label{eq:aggregate-throughput}
\Theta = \frac{p_{\mathrm{agg}}\,\Theta_T\,s_p}{t_s},
\end{equation}
where $\Theta$ denotes total throughput (bits/s), $p_{\mathrm{agg}}=\min(n_p,P_{\mathrm{Agg}}^{\mathrm{Trf}})$ is the average number of packets aggregated in a successful MU transmission, $s_p$ is the payload length (bits), and $\Theta_T=s_A + r_A$ is the expected number of successful MU transmissions per TF cycle (scheduled $s_A$ plus random-access $r_A$ contributions) \cite{behara2022performance}. The maximum aggregation per TF, $P_{\mathrm{Agg}}^{\mathrm{Trf}}$, is constrained by transmission opportunity (TXOP) duration ($t_d$), STA data rate, and PHY overhead:
\begin{equation}
P_{\mathrm{Agg}}^{\mathrm{Trf}} =
\min\!\Bigg(P_{\mathrm{Agg}}^{\mathrm{Max}},
\Big\lfloor \frac{T_r (t_d - T_{\mathrm{Phy}}^{\mathrm{MU}})}{s_t + s_h + s_p} \Big\rfloor \!\Bigg),
\end{equation}
with $P_{\mathrm{Agg}}^{\mathrm{Max}}$ the hardware/software aggregation limit, $T_r$ the STA rate, $s_t$ the Transport header size (bits), $s_h$ the Network header size (bits), and $s_p$ as in Equation \ref{eq:aggregate-throughput}, and $T_{\mathrm{Phy}}^{\mathrm{MU}}$ PHY overhead \cite{saldana2017frame,9353436}.

We obtain the per-STA transmission attempt probability $\beta_i$ from a Bianchi-style two-dimensional Markov chain extended for OFDMA/RU operation \cite{bianchi2002performance,behara2022performance,pokhrel2018modeling}. Under the decoupling assumption, the steady-state attempt probability is:
\begin{equation}
\label{eq:beta}
\beta_i = 
\frac{1 - \gamma^{mbo + fbo + 1}}
{(1 - \gamma)\displaystyle\sum_{k=0}^{mbo+fbo} B_k \gamma^k},
\end{equation}
where $\gamma$ is the conditional collision probability, $mbo$ is the maximum exponential backoff stage, $fbo$ the number of fixed backoff slots, and $B_k$ the average number of backoff slots at stage $k$. The coupled fixed-point equations for $\beta_i$ and $\gamma$ are solved iteratively as in \cite{behara2022performance}.

The collision probability due to random-access STAs is:
\begin{equation}
\label{eq:gamma_single}
\gamma = 1 - \left(1 - \frac{\beta_i}{r_A}\right)^{n_{\mathrm{ra}} - 1},
\end{equation}
where $n_{\mathrm{ra}}$ is the number of random-access STAs and $r_A$ is the random-access resource factor. A transmission attempt fails either due to a MAC collision or a PHY-layer error, giving the per-attempt failure probability:
\begin{equation}
\label{eq:fi}
f_i = 1 - (1 - \gamma)(1 - p_{\mathrm{phy}}),
\end{equation}
where $p_{\mathrm{phy}}$ is the PHY-layer packet error probability. Thus the per-attempt success probability is $(1-f_i)=(1-\gamma)(1-p_{\mathrm{phy}})$.

To convert event probabilities to time-domain quantities we model the expected slot duration seen by an arbitrary STA as \cite{pokhrel2018modeling}:
\begin{equation}
\label{eq:EX}
E[X] = P_0\sigma + P_sT_s + P_fT_c,
\end{equation}
where $P_0$, $P_s$, and $P_f$ are the probabilities of idle, successful, and failed slots, respectively, and $\sigma$, $T_s$, and $T_c$ are the corresponding slot durations. These probabilities depend on the attempt probabilities $\{\beta_j\}$ and $p_{\mathrm{phy}}$:

\[
\begin{aligned}
P_0 &= (1-\beta_{\mathrm{ap}})\!\prod_j(1-\beta_j)^{n_j},\\
P_s &= (1-\beta_{\mathrm{ap}})\!\left[1-\prod_j(1-\beta_j)^{n_j}(1-p_{\mathrm{phy}})\right],\\
P_f &= 1-P_0-P_s,
\end{aligned}
\] where $\beta_{\mathrm{ap}}$ is the AP attempt probability (if it contends) and $n_j$ is the number of STAs in class $j$. Following the MU-OFDMA timing in \cite{behara2022performance}, successful and collided slot durations are:

\[
\begin{aligned}
T_s &= t_s + \mathrm{DIFS},\\
T_c &= t_s - d_{\mathrm{mb}} + \mathrm{DIFS},
\end{aligned}
\] where $\mathrm{DIFS}=\mathrm{SIFS}+2\sigma$ under 802.11ax \cite{9353436}.

Finally, the effective throughput allocated to station $i$ is:
\begin{equation}
\label{eq:per_STA-throughput}
\Theta_i = \beta_i (1 - f_i)\,\Theta,
\end{equation}
which converts per-attempt success probabilities and aggregation into time-normalized service rates used by higher-layer models.

\subsection{BBRv3 Fluid Modeling}\label{sec:bbrv3_model}

Building on the per-STA service rate $\Theta_i$ from \eqref{eq:per_STA-throughput}, we adopt a fluid approximation of BBRv1 and BBRv2 from \cite{scherrer2022model} to BBRv3 to capture congestion-window evolution, pacing dynamics, and phase switching (\textit{ProbeBW} and \textit{ProbeRTT}) \cite{cardwell2023bbrv3,scherrer2022model}.

In the \textit{ProbeBW} phase the base congestion window $\overline w_i$ is decomposed into a high-bound component $w_i^{\mathrm{hi}}$ (probing) and a low-bound component $w_i^{\mathrm{lo}}$ (cruising). The effective \textit{ProbeBW} window is modeled as:
\begin{equation}
\label{eq:w_pbw}
w_i^{\mathrm{pbw}} =
\min \!\big( 2\overline w_i,\, m_i^{\mathrm{crs}} w_i^{\mathrm{lo}} \big)
+ \min \!\big( g_{\mathrm{hi}}\overline w_i,\, (1-m_i^{\mathrm{crs}}) w_i^{\mathrm{hi}} \big),
\end{equation}
where $g_{\mathrm{hi}}=2.25$ is the high-gain probe factor and $m_i^{\mathrm{crs}}\in[0,1]$ smoothly indicates the transition from probing ($\approx0$) to cruising ($\approx1$). The $\min(\cdot)$ operators bound in-flight to avoid instability under rapidly changing service rates.

The temporal evolution of the high and low-bound windows is driven by RTT feedback and loss-based multiplicative reduction:
\begin{align}
\dot{w}_i^{\mathrm{hi}} &= 
(1 - m_i^{\mathrm{crs}}) g_{\mathrm{hi}} 
\frac{t_i^{\mathrm{pbw}}}{\tau_i^{\min}}
\sigma(t_i^{\mathrm{pbw}} - \tau_i^{\min}) 
\sigma(v_i - w_i^{\mathrm{hi}}) \notag\\
&\quad - \frac{\Delta(p_{\pi_i})}{\tau_i^{\min}}
\sigma(p_{\pi_i} - p_{\mathrm{th}})\, w_i^{\mathrm{hi}}, 
\label{eq:dot_w_hi} \\
\dot{w}_i^{\mathrm{lo}} &= 
-(1 - m_i^{\mathrm{crs}}) \frac{1}{\tau_i^{\min}} 
(w_i^{\mathrm{lo}} - \overline w_i) \notag\\
&\quad - m_i^{\mathrm{crs}} \frac{\Delta(p_{\pi_i})}{\tau_i^{\min}}
\sigma(p_{\pi_i} - p_{\mathrm{th}})\, w_i^{\mathrm{lo}}.
\label{eq:dot_w_lo}
\end{align}

In the above, $\tau_i^{\min}$ denotes the minimum RTT (propagation delay), $t_i^{\mathrm{pbw}}$ is the RTT sample taken during \textit{ProbeBW}, $v_i$ is instantaneous in-flight, and $\sigma(\cdot)$ is a smooth activation (e.g., sigmoid) used to approximate discrete state transitions. The function $\Delta(p_{\pi_i})$ maps instantaneous packet-loss probability $p_{\pi_i}$ to the multiplicative reduction intensity; $p_{\mathrm{th}}$ is the loss threshold that triggers a reduction.

The instantaneous sending rate during \textit{ProbeBW} is then:
\begin{equation}
\label{eq:x_pbw}
x_i^{\mathrm{pbw}}(t) = \frac{w_i^{\mathrm{pbw}}}{\tau_i^{\min}}.
\end{equation}

During \textit{ProbeRTT}, BBRv3 reduces in-flight to re-estimate minimum RTT by draining queues \cite{cardwell2023bbrv3}:
\begin{equation}
\label{eq:prt}
w_i^{\mathrm{prt}} = \frac{\overline w_i}{2}, 
\qquad 
x_i^{\mathrm{prt}}(t) = \frac{\overline w_i}{2 \tau_i^{\min}}.
\end{equation}
The active sending rate alternates between \textit{ProbeRTT} and \textit{ProbeBW} according to protocol state:
\begin{equation}
\label{eq:x_switch}
x_i(t) =
\begin{cases}
x_i^{\mathrm{prt}}(t), & \text{if$ ProbeRTT$ active},\\[2pt]
x_i^{\mathrm{pbw}}(t), & \text{if \textit{ProbeBW} active}.
\end{cases}
\end{equation}

Together, \eqref{eq:w_pbw}--\eqref{eq:x_switch} complete the BBRv3 transport-layer behavior, which we now couple with MAC-layer service in the cross-layer RTT and queue dynamics.

\subsection{Cross-Layer RTT and Queue Dynamics}
\label{sec:cross_layer}

The MAC-layer service $\Theta_i$ and transport sending rate $x_i(t)$ couple through queue occupancy at the AP. We model the observed RTT for flow $i$ as:
\begin{equation}
\label{eq:total_rtt}
\tau_i(t) = \tau_i^{\min} + \frac{q_i(t)}{\Theta_i},
\end{equation}
where $q_i(t)$ is the instantaneous per-flow queue occupancy and $\tfrac{q_i(t)}{\Theta_i}$ is the fluid-approximation queuing delay (Little's Law) \cite{little1961proof}.

The queue dynamics follow the standard fluid TCP/AQM expression:
\begin{equation}
\label{eq:queue_dynamics}
\frac{d q_i(t)}{d t} = x_i(t) - \Theta_i,
\end{equation}
i.e., backlog increases when sending exceeds service and drains otherwise. In the presence of AQM-induced drops/marks, the effective throughput received by flow $i$ is:
\begin{equation}
\label{eq:effective_throughput}
\Theta_i(t) = (1 - p_i(t))\, x_i(t),
\end{equation}
where $p_i(t)$ is the instantaneous packet drop/marking probability at the queue or MAC level. Equations \eqref{eq:total_rtt}--\eqref{eq:effective_throughput} close the feedback loop between BBRv3's estimators and the MAC/AQM behaviors.

\subsection{Active Queue Management (AQM) Modeling}
\label{sec:aqm_model}

We incorporate representative AQM schemes that operate at the bottleneck queue and determine $p_i(t)$ in \eqref{eq:effective_throughput}. These AQM models are used both in analysis and to interpret experimental outcomes.

\medskip
\noindent\textbf{1) FIFO (PFIFO / DropTail).}  
Under FIFO, all stations share a common queue of capacity $Q_{\max}$. The instantaneous per-station service allocation (used for modeling aggregate effects) is:
\begin{align}
s_i^{\mathrm{fifo}}(t) &= 
\frac{\Theta_i}{\sum_j \Theta_j}\,
\min\!\Big\{\sum_j q_j(t),\,Q_{\max}\Big\},
\label{eq:si_fifo}
\\[2pt]
d_i^{\mathrm{fifo}}(t) &=
p_{\mathrm{drop}}\!\big(q_{\mathrm{tot}}(t)\big)\,x_i(t),
\qquad 
q_{\mathrm{tot}}(t)=\sum_j q_j(t),
\label{eq:di_fifo}
\end{align}
where $s_i^{\mathrm{fifo}}(t)$ represents the FIFO share based on nominal weights $\Theta_i$,  
$d_i^{\mathrm{fifo}}(t)$ models drop-tail/RED-style loss,  
and $p_{\mathrm{drop}}(\cdot)$ is either an indicator or a linear RED-like function across $[Q_{\min},Q_{\max}]$ \cite{floyd2008internet}.

\medskip
\noindent\textbf{2) FQ-CoDel.}  
FQ-CoDel combines per-flow queuing (WFQ/DRR approximation of GPS) with CoDel delay-based dropping \cite{hoeiland2018flow}. In fluid form:
\begin{align}
s_i^{\mathrm{fq}}(t) &= \alpha_i^{\mathrm{fq}}(t)\,\Theta, 
\qquad 
\alpha_i^{\mathrm{fq}}(t) = \frac{w_i^{\mathrm{fq}}(t)}{\sum_j w_j^{\mathrm{fq}}(t)},
\label{eq:si_fq}
\\[2pt]
\tau_i^{\mathrm{soj}}(t) &= \frac{q_i(t)}{s_i^{\mathrm{fq}}(t)}, 
\qquad 
d_i^{\mathrm{fq}}(t) = \mathbf{1}\{\tau_i^{\mathrm{soj}}(t) > \mathrm{target}\}\cdot \kappa_{\mathrm{codel}},
\label{eq:drop_fq}
\end{align}
where $w_i^{\mathrm{fq}}(t)$ denotes FQ-CoDel’s per-flow scheduling weight,  
$\alpha_i^{\mathrm{fq}}(t)$ is the normalized weight determining the share of $\Theta$,  
$\tau_i^{\mathrm{soj}}(t)$ is the sojourn (queuing) time, and  
$d_i^{\mathrm{fq}}(t)$ applies CoDel’s drop/mark behavior when delay persists beyond the target threshold.

\medskip
\noindent\textbf{3) CAKE.}  
CAKE extends FQ-CoDel with per-host fairness and an adaptive marking function. In fluid form:
\begin{align}
\alpha_i^{\mathrm{cake}}(t) &= 
\frac{\omega_i(t)}{\sum_j \omega_j(t)}, 
\qquad 
\omega_i(t) = \phi_{\mathrm{host}}\!\big(h(i)\big)\cdot \psi_i,
\label{eq:alpha_cake}
\\[2pt]
s_i^{\mathrm{cake}}(t) &= \alpha_i^{\mathrm{cake}}(t)\,\Theta, 
\qquad 
m_i^{\mathrm{cake}}(t) = F_{\mathrm{cake}}\!\big(\tau_i^{\mathrm{soj}}(t),\,q_i(t)\big),
\label{eq:si_cake}
\end{align}
where $\phi_{\mathrm{host}}(h(i))$ applies CAKE’s per-host fairness,  
$\psi_i$ is the per-flow fairness weight,  
$\alpha_i^{\mathrm{cake}}(t)$ is the normalized CAKE service share, and  
$m_i^{\mathrm{cake}}(t)$ is CAKE’s nonlinear drop/mark probability as a function of sojourn time and backlog \cite{hoiland2018piece,nichols2012controlling,pan2013pie}.  
The queue evolution under CAKE is:
\begin{equation}
\frac{dq_i(t)}{dt} = x_i(t)\big(1 - m_i^{\mathrm{cake}}(t)\big) - s_i^{\mathrm{cake}}(t),
\label{eq:cake_dynamics}
\end{equation}
where $x_i(t)$ is the offered load and the term $1 - m_i^{\mathrm{cake}}(t)$ captures CAKE’s adaptive ECN/drop behavior.

\begin{figure} [t]
    \centering
   \includegraphics[scale=0.18]{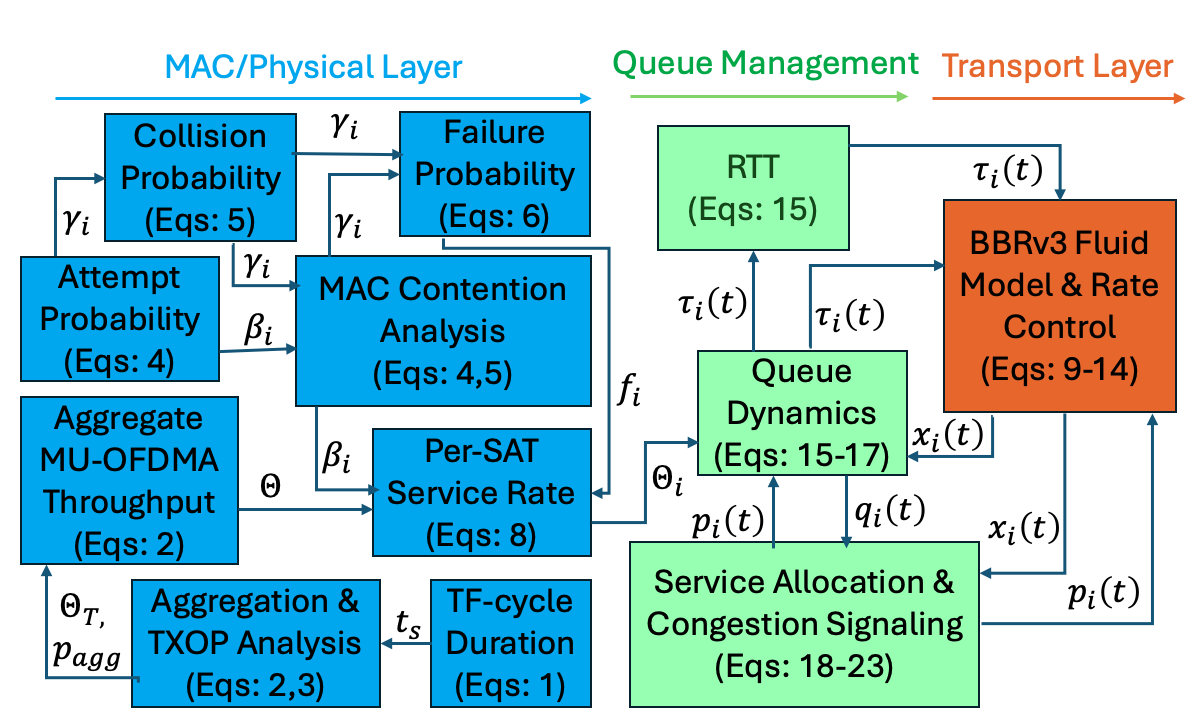}
    \caption{Details of the modules of the Cross-layer analytical framework integrating MU–OFDMA-based MAC throughput
modelling, queue dynamics, congestion signaling, and BBRv3 packet dynamics control.}
\label{fig:modeling}
\end{figure}

The Figure~\ref{fig:modeling} demonstrates the complete cross-layer performance interaction of a BBRv3 flow operating over IEEE~802.11ax MU–OFDMA with a modern AQM system. Observe in Figure~\ref{fig:modeling} that the MAC/Physical Layer (Eqs.~1--8) derives the per-STA service rate $\Theta_i$ using TF-cycle timing, aggregation probability, OFDMA throughput, contention behaviour, and failure probability. This service rate drives the Queue Dynamics module (Eqs.~15--17), which models queue evolution $q_i(t)$, effective throughput, and round-trip time $\tau_i(t)$. The RTT feeds the Transport-Layer BBRv3 Fluid Model (Eqs.~9--14) to determine the sending rate $x_i(t)$, which in turn contributes to queue buildup. The Service Allocation and Congestion Signaling module (Eqs.~18--23), representing PFIFO, FQ-CoDel, and CAKE queue disciplines, derives congestion feedback $p_i(t)$ and flow-based service shares from $q_i(t)$ and $x_i(t)$. The resulting feedback $p_i(t)$ affects both Queue Dynamics and BBRv3, forming a closed-loop coupling across MAC, queue, and transport layers. This model captures the mutual influence between OFDMA scheduling, queuing behaviour, and BBRv3 congestion control in Wi-Fi networks.

\section{Experimental Testbed} \label{experimentaltestbed}

All experiments were conducted on a custom-designed wireless testbed replicating modern residential network dynamics. The setup was built to evaluate TCP congestion control performance under controlled yet realistic Wi-Fi conditions representative of smart-home environments. Key objectives included precise bandwidth control, isolation of queuing disciplines, and reproducible bidirectional TCP flows. The testbed supports fine-grained traffic measurement, enabling detailed analysis of congestion control algorithms under varied network conditions. All experiments were repeated five times for experimental repeatability.

\begin{figure} [h]
    \centering
    \includegraphics[scale=0.25]{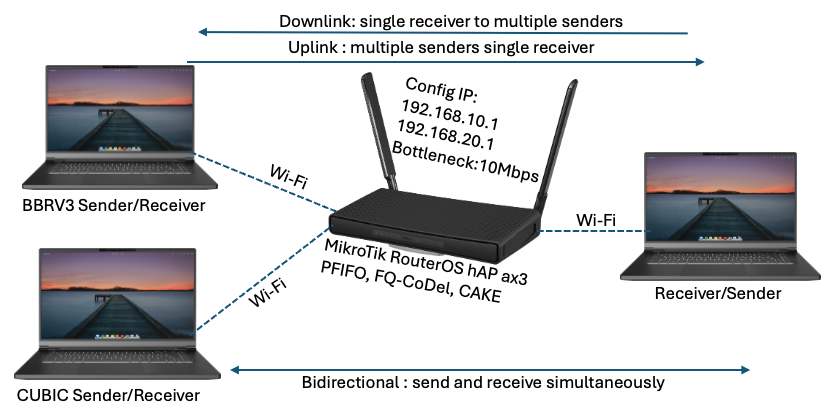}
    \caption{Experimental testbed setup.}
    \label{fig:testbed}
\end{figure}

\subsection{Wireless Testbed Architecture}
\label{sec:testbed}

Figure~\ref{fig:testbed} illustrates the custom testbed with a MikroTik RouterOS hAP ax3 router serving as the central AP, managing client connections and applying AQM policies. Three dedicated laptop nodes act as traffic generators and receivers. The BBRv3 Sender/Receiver node generates TCP flows using the BBRv3 CCA, while the CUBIC Sender/Receiver node uses CUBIC. All devices connect to the AP over Wi-Fi, creating a fully wireless environment typical in smart homes.

The blue double-headed arrows in Figure~\ref{fig:testbed} denote bidirectional traffic patterns. The top arrow shows the \emph{Downlink: single receiver to multiple senders} case, where the Receiver/Sender node collects data from two senders (one CUBIC, one BBRv3). Conversely, \emph{Uplink: multiple senders to single receiver} refers to traffic originating from both sender nodes and terminating at the Receiver/Sender. The bottom arrow highlights simultaneous send/receive capability, enabling evaluation of concurrent bidirectional TCP flows, characteristic of interactive home applications.

The testbed uses two logical subnets for client and server traffic: 192.168.10.0/24 and 192.168.20.0/24. The MikroTik router (RouterOS v7.8) enforces a 10~Mbps bandwidth limit in both directions via queue tree mechanisms and packet-mark-based classification, creating the bottleneck link. Queuing disciplines are applied per flow using static policy assignment: PFIFO (50-packet buffer) to study drop-tail behavior, and CAKE and FQ-CoDel with default AQM configurations in the router. These settings allow in-depth analysis of CCA behaviors under contention and queuing-induced bottlenecks.

\subsection{Traffic Patterns and Queue Disciplines}

The study evaluates BBRv3 performance in multi-device Wi-Fi environments across three traffic patterns:
\textbf{Uplink (UL)}: multiple devices transmitting to a single endpoint
\textbf{Downlink (DL)}: single source transmitting to multiple recipients
\textbf{Bidirectional simultaneous transmission:} concurrent data exchange between endpoints.

Comparisons are made between two CCAs: CUBIC and BBRv3 across varying endpoint configurations under two queue management setups:\textbf{Baseline}: FIFO drop-tail queuing without active management and, \textbf{Advanced AQM}: FQ-CoDel and CAKE.

% These AQM algorithms are designed to improve queue management, sustain consistent performance, and ensure fair resource allocation across competing flows~\cite{hoeiland2018flow,hoiland2018piece,nichols2018controlled}.

\subsection{Measurement and Analysis Tools}

TCP flows are generated with iperf3 (v3.9). Socket-level statistics are collected in real time using \texttt{ss -tin}, and packet-level traces are recorded via TShark (v4.4.6). Outputs are parsed using cJSON (v1.7.3) into structured JSON and plain text logs. Collected metrics include throughput, $cwnd$, RTT, jitter, and retransmissions, which enables reproducible analysis of pacing delivery dynamics and TCP performance under diverse queue regimes. All sender/receiver nodes run Ubuntu 22.04.5 LTS for a consistent experimental environment.

\section{Evaluation and Analysis} \label{analysis}

This section presents a comprehensive evaluation of BBRv3 and CUBIC across three queue disciplines PFIFO, FQ-CoDel, and CAKE within a Wi-Fi environment. The analysis is structured around BBR’s advancements, its pacing and delivery rate dynamics, and the fairness, responsiveness, and coexistence behavior of BBRv3 and CUBIC under contention in uplink, downlink, and bidirectional scenarios. These results are obtained experimentally and are supported by the analytical cross-layer model presented in Section~\ref{sec:system_model}.

\subsection{BBR Advancement}

\begin{figure}[htbp]
    \centering
    \begin{subfigure}[b]{0.15\textwidth}
        \centering
        \includegraphics[width=\textwidth]{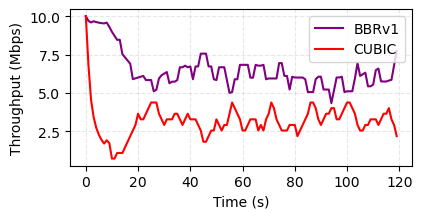}
        \caption{}
        \label{fig:bbr1_cubic}
    \end{subfigure}
    \hfill
    \begin{subfigure}[b]{0.15\textwidth}
        \centering
        \includegraphics[width=\textwidth]{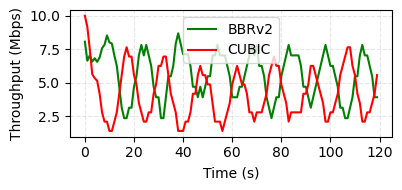}
        \caption{}
        \label{fig:bbr2_cubic}
    \end{subfigure}
    \hfill
    \begin{subfigure}[b]{0.15\textwidth}
        \centering
        \includegraphics[width=\textwidth]{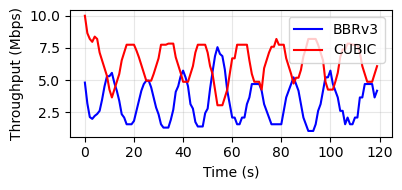}
        \caption{}
        \label{fig:bbr3_cubic}
    \end{subfigure}

    \caption{Throughput comparison of BBRv1, BBRv2, and BBRv3 competing with CUBIC over Wi-Fi.}
    \label{bbr_advancement}
\end{figure}

The throughput results in Figure~\ref{bbr_advancement} compare CUBIC with BBRv1, BBRv2, and BBRv3 under identical Wi-Fi conditions. In particular, the MAC-layer service rate $\Theta_i$ in Eq.~\ref{eq:per_STA-throughput} shapes the available capacity to each station, while the RTT queue coupling captured in Eqs.~\ref{eq:total_rtt}--\ref{eq:queue_dynamics} governs delay build-up and feedback timing. Together with BBRv3's pacing and window evolution rules (Eqs.~\ref{eq:w_pbw}--\ref{eq:x_switch}), these components account for the throughput differences observed across the CCAs.

\smallskip
\noindent\textbf{BBRv1 vs. CUBIC.}
Figure~\ref{fig:bbr1_cubic} shows that BBRv1 (purple) consistently dominates CUBIC (red), frequently exceeding 7.5\,Mbps while CUBIC struggles below 5\,Mbps. This behavior follows directly from BBRv1's tendency to sustain a sending rate above the available service $\Theta_i$, which according to the queue evolution in Eq.~\ref{eq:queue_dynamics} induces persistent queue buildup. The resulting RTT inflation predicted by Eq.~\ref{eq:total_rtt} further suppresses CUBIC, which relies on packet loss rather than delay to detect congestion. Wi-Fi’s variable and contention-driven losses exacerbate this imbalance, producing the experimentally observed unfairness and near starvation of CUBIC.

\smallskip
\noindent\textbf{BBRv2 vs. CUBIC.}
Figure~\ref{fig:bbr2_cubic} demonstrates the opposite extreme: BBRv2 (green) frequently yields bandwidth to CUBIC, sometimes falling below 2\,Mbps. This is also consistent with the analytical model. BBRv2 responds aggressively to increases in instantaneous loss probability $p_{\pi_i}$, reducing its in-flight as dictated by the loss-driven terms in Eqs.~\ref{eq:dot_w_hi}--\ref{eq:dot_w_lo}. Because many Wi-Fi losses are not congestion-driven, BBRv2 reduces its rate more often than required, allowing CUBIC’s cubic growth function to occupy the freed capacity. The result is improved coexistence relative to BBRv1, but at the cost of chronic under-utilization.

\smallskip
\noindent\textbf{BBRv3 vs. CUBIC.}
Figure~\ref{fig:bbr3_cubic} reveals a more intricate pattern: BBRv3 (blue) and CUBIC (red) exhibit large, inverse oscillations, where one surges toward 7.5\,Mbps while the other collapses toward 2–3\,Mbps. This oscillatory structure is consistent with the interaction between BBRv3’s periodic \textit{ProbeBW} pacing adjustments (modeled in Eq.~\ref{eq:w_pbw}) and the time-varying Wi-Fi service $\Theta_i(t)$. When BBRv3 temporarily overshoots $\Theta_i(t)$, queue buildup (Eq.~\ref{eq:queue_dynamics}) triggers RTT inflation (Eq.~\ref{eq:total_rtt}), prompting BBRv3 to reduce its pacing rate and enabling CUBIC to reclaim bandwidth. Because $\Theta_i(t)$ fluctuates due to contention, aggregation, and backoff, this feedback loop manifests as a persistent limit cycle rather than a stable sharing point. A qualitative oscillation rate $\omega$ can be interpreted as the number of throughput swings per unit time.

% \smallskip
% \noindent\textbf{Summary.}
% Across versions, the experimental patterns reflect the mechanisms captured by the analytical model:  
% BBRv1 overshoots and starves CUBIC;  
% BBRv2 overreacts to Wi-Fi loss and yields excessively;  
% BBRv3 improves coexistence but remains susceptible to cross-layer feedback, failing to converge to a stable equilibrium without queue assistance.

\textit{\textbf{Takeaway:}  
BBRv3 mitigates the extremes of earlier versions but still oscillates under Wi-Fi’s variable service rate. These oscillations arise from the RTT–queue feedback loop and the pacing dynamics formalized in the analytical model, highlighting the need for AQM support for stable coexistence.}

% ----------------------------------------------------------------------

\subsection{BBRv3 Pacing--Delivery Dynamics and AQM Impact}

\begin{figure}[htbp]
    \centering
    \begin{subfigure}[b]{0.15\textwidth}
        \centering
        \includegraphics[width=\textwidth]{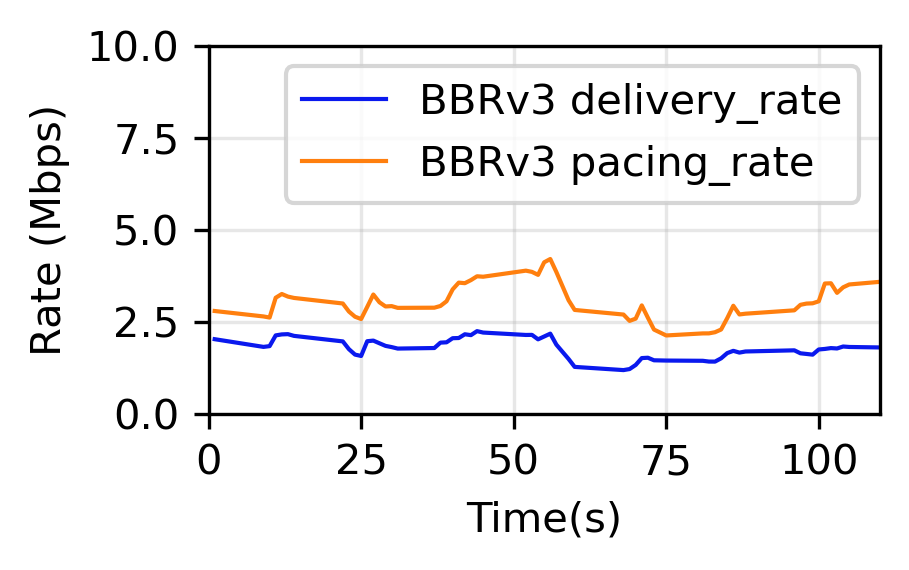}
        \caption{PFIFO}
        \label{fig:fifo_pacing}
    \end{subfigure}
    \hfill
    \begin{subfigure}[b]{0.15\textwidth}
        \centering
        \includegraphics[width=\textwidth]{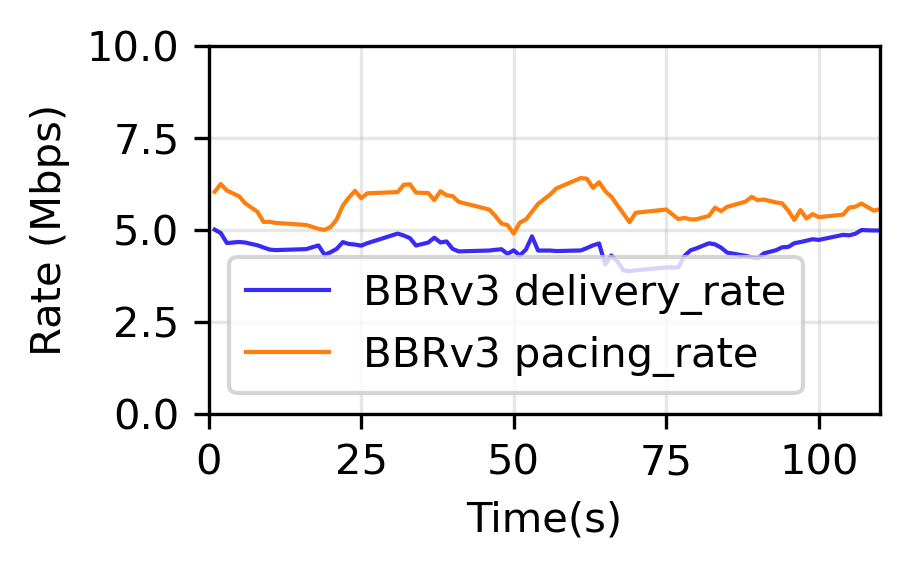}
        \caption{FQ-CoDel}
        \label{fig:fqcodel_pacing}
    \end{subfigure}
    \hfill
    \begin{subfigure}[b]{0.15\textwidth}
        \centering
        \includegraphics[width=\textwidth]{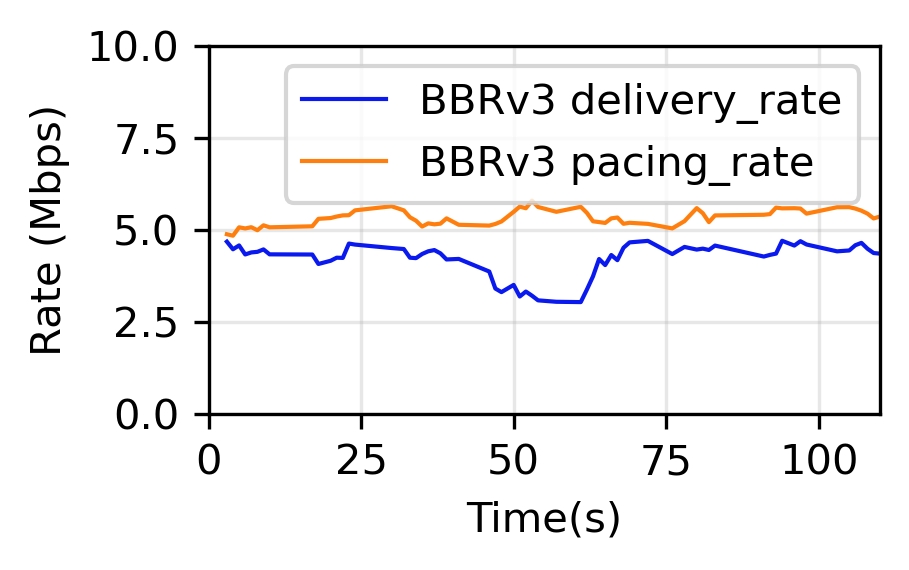}
        \caption{CAKE}
        \label{fig:cake_pacing}
    \end{subfigure}
 
    \caption{Pacing and delivery rate evolution of BBRv3 under PFIFO, FQ-CoDel, and CAKE when competing with a CUBIC flow over a 10\,Mbps Wi-Fi bottleneck.}
    \label{fig:pacing_rate}
\end{figure}

Figure~\ref{fig:pacing_rate} illustrates BBRv3’s pacing and delivery rate evolution while competing with CUBIC under three queue disciplines. These traces expose how well BBRv3’s internal model represented by its pacing decisions in Eq.~\ref{eq:w_pbw} aligns with the actual service delivered to the flow. The degree of alignment reflects whether BBRv3’s rate-based control is respected or distorted by the underlying queue.

\smallskip

Under PFIFO (Figure~\ref{fig:fifo_pacing}), pacing stabilizes around 5--6\,Mbps, but delivery remains below 2\,Mbps. This persistent mismatch indicates that PFIFO’s burst-sensitive behavior disrupts BBRv3’s probing, causing queue build-up and losses inconsistent with its model assumptions. The queuing dynamics in Eq.~\ref{eq:queue_dynamics} predict such divergence when bursts from competing flows dominate queue occupancy.

\smallskip

With FQ-CoDel (Figure~\ref{fig:fqcodel_pacing}), pacing occasionally aligns with delivery, producing intermittent peaks around 3--4.5\,Mbps. Although the AQM partially isolates flows and controls queue delay, its per-packet fairness and drop decisions remain insufficiently aligned with BBRv3’s pacing cycles, resulting in volatile behavior. This is consistent with the RTT feedback mechanism formalized in Eqs.~\ref{eq:total_rtt}--\ref{eq:queue_dynamics}.

\smallskip

CAKE (Figure~\ref{fig:cake_pacing}) presents a markedly different scenario: pacing around 5.5\,Mbps closely matches delivery near 4.5\,Mbps. This stable relationship reflects CAKE’s per-flow fairness and rate shaping, which maintain predictable queue occupancy and consistent RTT trends. As a result, BBRv3’s probing behavior (governed by Eq.~\ref{eq:w_pbw}), directly translates into actual throughput, enabling stable coexistence with CUBIC. These results align with observations in~\cite{shrestha2025visualizing}.

% \smallskip
% \noindent\textbf{Summary.}
% Figure~\ref{fig:pacing_rate} shows that BBRv3’s ability to realize its internal model depends strongly on the queue discipline. Without fairness-oriented AQM, pacing accuracy does not translate into delivery, leading to underutilization or instability.

\textit{\textbf{Takeaway:}  
BBRv3's pacing model operates reliably only when supported by queue-aware AQM (e.g., CAKE). PFIFO and, to a lesser extent, FQ-CoDel distort the RTT–queue feedback loop fundamental to the design of BBRv3, weakening the alignment between pace and delivery and affecting coexistence.}

\subsection{Uplink Scenario}

We now examine the uplink behavior of CUBIC and BBRv3 when multiple clients transmit simultaneously over a shared Wi-Fi bottleneck. The MAC-layer service rate $\Theta_i$ (Eq.~\ref{eq:per_STA-throughput}), queue dynamics (Eq.~\ref{eq:queue_dynamics}), and RTT coupling (Eq.~\ref{eq:total_rtt}) collectively determine how each CCA reacts to the time-varying contention and loss conditions, as illustrated in the experimental graphs.

\begin{figure}[htbp]
    \centering
    \begin{subfigure}[b]{0.15\textwidth}
        \centering
         \includegraphics[width=\textwidth]{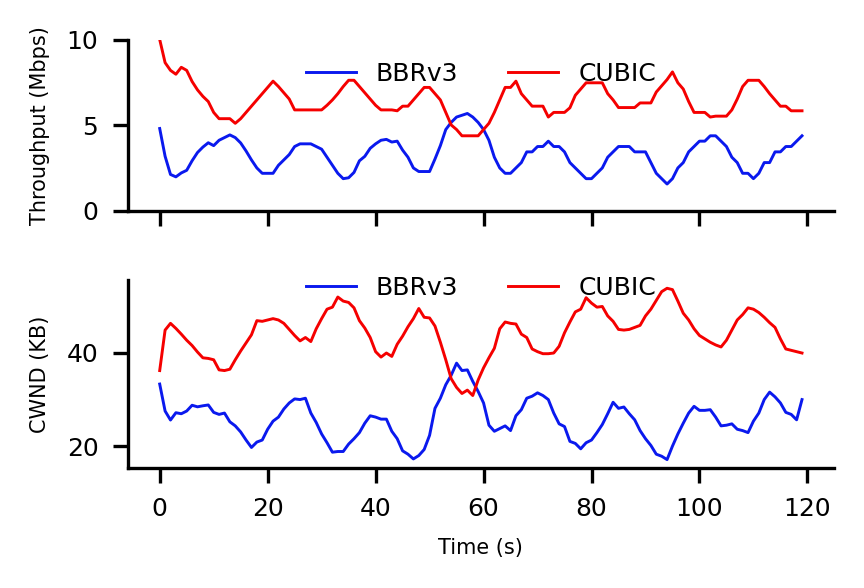}
        \caption{PFIFO}
        \label{fig:pfifo_thpt_up}
    \end{subfigure}
    \hfill
    \begin{subfigure}[b]{0.15\textwidth}
        \centering
        \includegraphics[width=\textwidth]{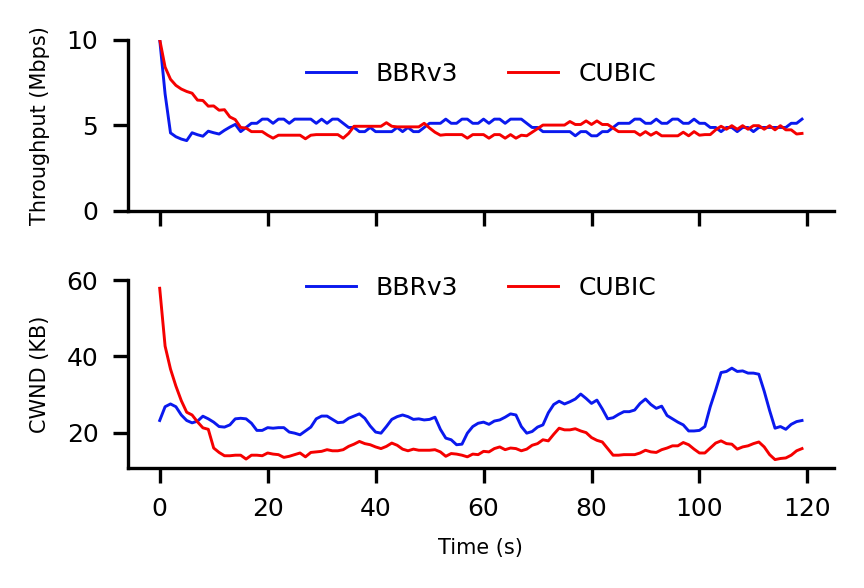}
        \caption{FQ-CoDel}
        \label{fig:fq-codel_thpt_up}
    \end{subfigure}
    \hfill
    \begin{subfigure}[b]{0.15\textwidth}
        \centering
        \includegraphics[width=\textwidth]{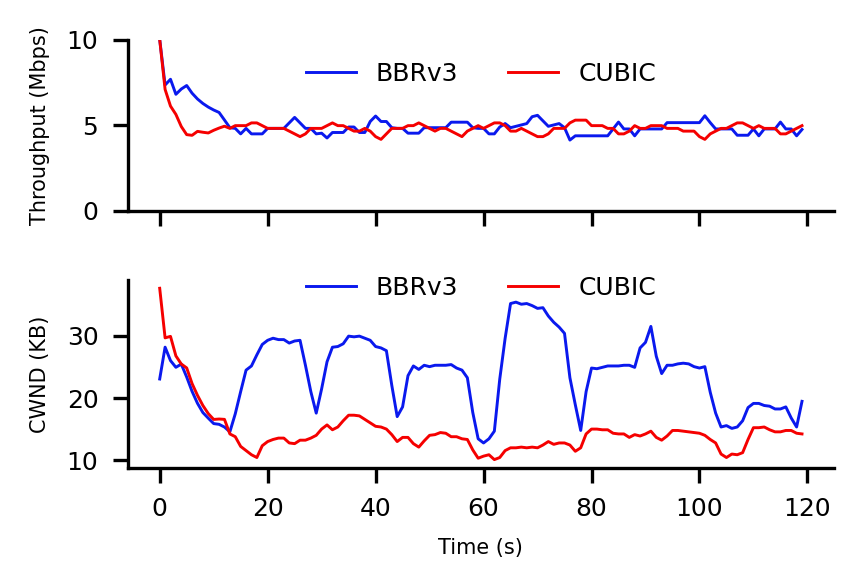}
        \caption{CAKE}
        \label{fig:cake_thpt_up}
    \end{subfigure}
 
    \caption{Uplink throughput (top) and $cwnd$ (bottom) dynamics for CUBIC (red) and BBRv3 (blue) under PFIFO, FQ-CoDel, and CAKE over a 10 Mbps Wi-Fi link.}
    \label{fig:allthroughput_up}
\end{figure}

Under PFIFO (Figure~\ref{fig:pfifo_thpt_up}), CUBIC obtains a disproportionate share of the link, fluctuating between 6--8\,Mbps, while BBRv3 remains restricted to 2--4\,Mbps. This strong imbalance is predicted by the analytical queue model: without per-flow isolation, CUBIC’s additive-increase behavior drives its in-flight persistently above the service allocation $s_i^{\mathrm{fifo}}(t)$ (Eq.~\ref{eq:si_fifo}), causing queue accumulation as described by Eq.~\ref{eq:queue_dynamics}. The resulting RTT inflation (Eq.~\ref{eq:total_rtt}) suppresses BBRv3’s ability to maintain an accurate estimate of bottleneck bandwidth and \texttt{minRTT}, making its pacing decisions (Eq.~\ref{eq:w_pbw}) overly conservative. The $cwnd$ traces reflect this: CUBIC oscillates with large saw-tooth swings ($\sim$35--50\,KB), whereas BBRv3 remains in a narrower but still unstable 20--30\,KB range. PFIFO therefore amplifies asymmetry between loss-driven and model-driven CCAs, leading to persistent uplink unfairness.

With FQ-CoDel (Figure~\ref{fig:fq-codel_thpt_up}), throughput becomes significantly more balanced. Both flows stabilize near 5\,Mbps with reduced oscillatory behavior. This aligns with the analytical expectation that isolating each flow into independent virtual queues effectively equalizes the service allocation $s_i^{\mathrm{fq}}(t)$ (Eq.~\ref{eq:si_fq}) for each sender. CoDel's delay-based signaling keeps queue occupancy bounded, limiting RTT excursions predicted by Eq.~\ref{eq:total_rtt}. For CUBIC, earlier congestion signals prevent buffer saturation; for BBRv3, tighter RTT distributions improve the fidelity of its bandwidth probing cycle (Eq.~\ref{eq:w_pbw}), avoiding starvation. Accordingly, $cwnd$ traces remain contained: CUBIC between 15--20\,KB, BBRv3 between 20--30\,KB. These behaviors reflect FQ-CoDel’s capacity to realign practical flow dynamics with the theoretical per-flow service model.

CAKE (Figure~\ref{fig:cake_thpt_up}) provides the most stable and equitable sharing, with both flows sustaining $\approx$5\,Mbps with minimal variability. Beyond per-flow fairness, CAKE’s per-host fairness and enforced shaping ensure that the effective service $\Theta_i$ delivered to each sender remains tightly bounded. This produces RTT and queue behavior highly consistent with the analytical queue model (Eq.~\ref{eq:queue_dynamics}) and results in near-perfect pacing delivery alignment. BBRv3 occasionally increases its $cwnd$ to $\sim$35\,KB when probing, as predicted by the \textit{ProbeBW} cycle in Eq.~\ref{eq:w_pbw}, but CAKE's shaping and adaptive drop probability $m_i^{\mathrm{cake}}(t)$ (Eq.~\ref{eq:si_cake}) restricts the actual transmission rate, preventing bursts from disrupting coexistence. CUBIC’s $cwnd$ remains tightly between 12--18\,KB due to responsive congestion signaling. Overall, CAKE delivers an uplink environment where both loss-driven and model-driven CCAs converge toward the analytically expected steady state.

\begin{figure}[htbp]
    \centering
    \begin{subfigure}[b]{0.15\textwidth}
        \centering
        \includegraphics[width=\textwidth]{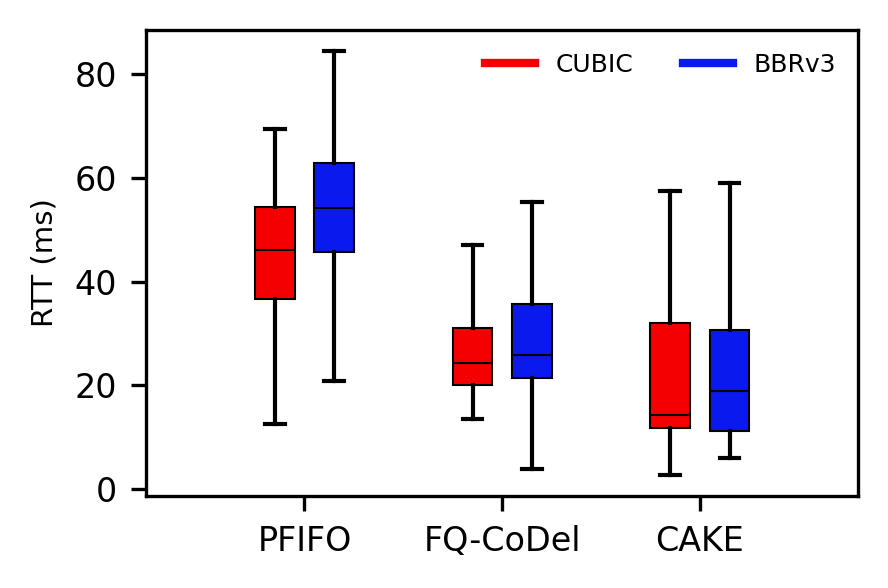}
        \caption{UL RTT}
        \label{fig:rtt_up}
    \end{subfigure}
    \hfill
    \begin{subfigure}[b]{0.15\textwidth}
        \centering
        \includegraphics[width=\textwidth]{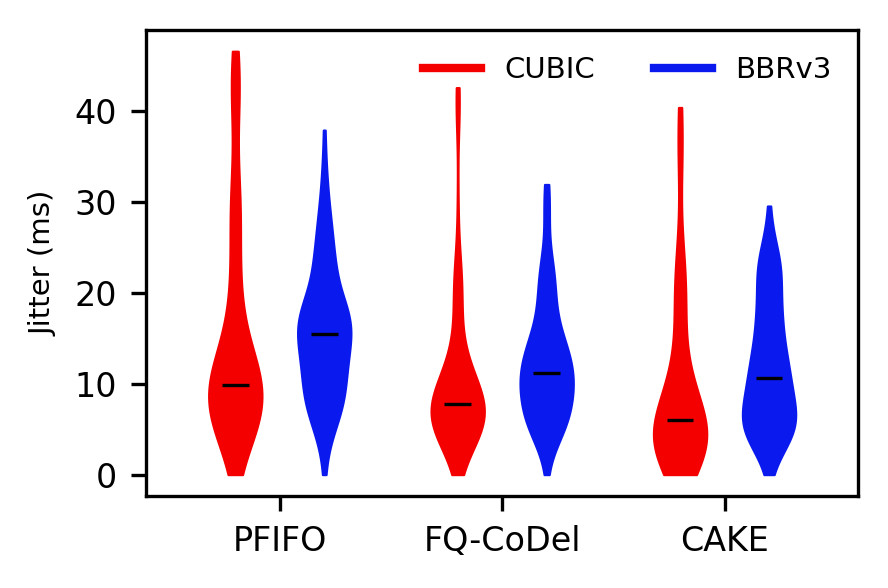}
        \caption{UL Jitter}
        \label{fig:jitter_up}
    \end{subfigure}
    \hfill
    \begin{subfigure}[b]{0.17\textwidth}
        \centering
        \includegraphics[width=\textwidth]{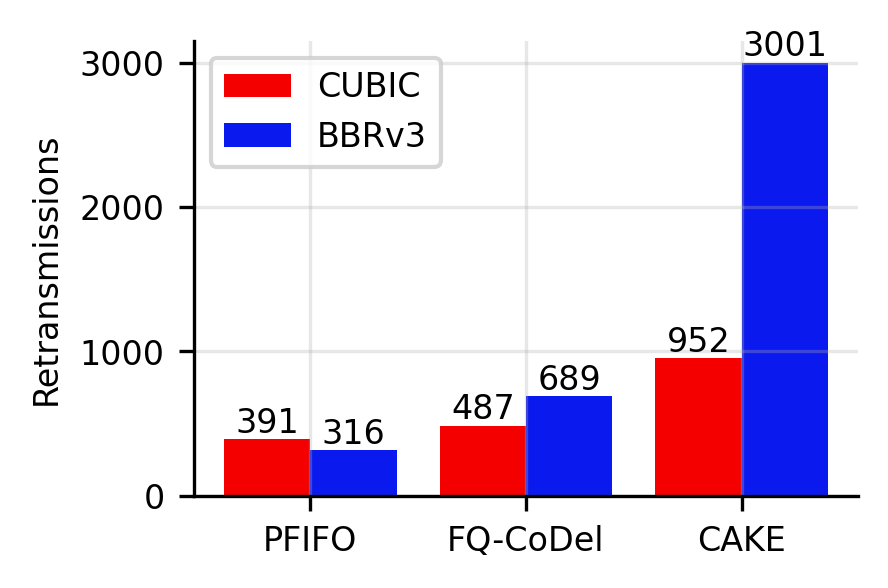}
        \caption{UL Retransmission}
        \label{fig:retransmission_up}
    \end{subfigure}
 
    \caption{RTT, jitter, and retransmission statistics for uplink CUBIC and BBRv3 across AQMs.}
    \label{fig:3in1}
\end{figure}

Figure~\ref{fig:rtt_up} shows that PFIFO produces the highest median RTT (50--55\,ms) and the widest spread, consistent with the queue accumulation predicted by Eq.~\ref{eq:queue_dynamics}. FQ-CoDel reduces RTTs to $\sim$25\,ms by actively controlling queue length, while CAKE achieves the lowest median RTT ($\sim$18\,ms) with the tightest distribution. These stable RTTs directly benefit BBRv3, whose bandwidth estimation and pacing decisions depend on accurate RTT sampling (Eq.~\ref{eq:total_rtt}).

Figure~\ref{fig:jitter_up} reveals similar trends. PFIFO yields the highest jitter ($\sim$14\,ms median) due to uncontrolled queue growth, whereas FQ-CoDel and CAKE achieve substantially lower jitter through active delay control. Lower jitter improves BBRv3’s model accuracy and reduces $cwnd$ overreaction, aligning practical dynamics with the analytical pacing behavior.

Figure~\ref{fig:retransmission_up} shows retransmission counts across AQMs. PFIFO generates moderate retransmissions for both flows due to buffer overflows. FQ-CoDel increases retransmissions slightly as early drops provide timely congestion cues. Under CAKE, BBRv3 experiences significantly higher retransmissions than CUBIC (3001 vs.\ 952). This arises from CAKE's adaptive drop probability $m_i^{\mathrm{cake}}(t)$ (Eq.~\ref{eq:si_cake}), where BBRv3’s aggressive probing leads to more drops, amplifying retransmissions. In contrast, CUBIC reduces its window after each loss, avoiding persistent probing and thus incurring fewer retransmissions.

\textit{\textbf{Takeaway:}  
In Wi-Fi uplinks, PFIFO strongly favors aggressive CCAs such as CUBIC, exacerbating RTT inflation and starving BBRv3. FQ-CoDel restores fairness by ensuring predictable per-flow service and stable delay, while CAKE provides the most model-aligned behavior with consistent RTTs, low jitter, and balanced throughput. CAKE’s tight shaping and fairness controls allow BBRv3’s analytical pacing model to manifest accurately in practice, making it the most effective AQM for mixed-CCA uplink scenarios.}

\subsection{Downlink Scenario}

We now evaluate the downlink case, where a single AP sends data to multiple stations. In the downlink, the AP is the sole sender, so each flow experiences a more regular MAC-layer service rate $\Theta_i$ (Eq.~\ref{eq:per_STA-throughput}) and less severe contention than in uplink transmissions. However, queue dynamics (Eq.~\ref{eq:queue_dynamics}) and RTT coupling (Eq.~\ref{eq:total_rtt}) continue to differentiate PFIFO, FQ-CoDel, and CAKE in terms of fairness, delay, and retransmission behavior.

\begin{figure}[htbp]
    \centering
    \begin{subfigure}[b]{0.15\textwidth}
        \centering
         \includegraphics[width=\textwidth]{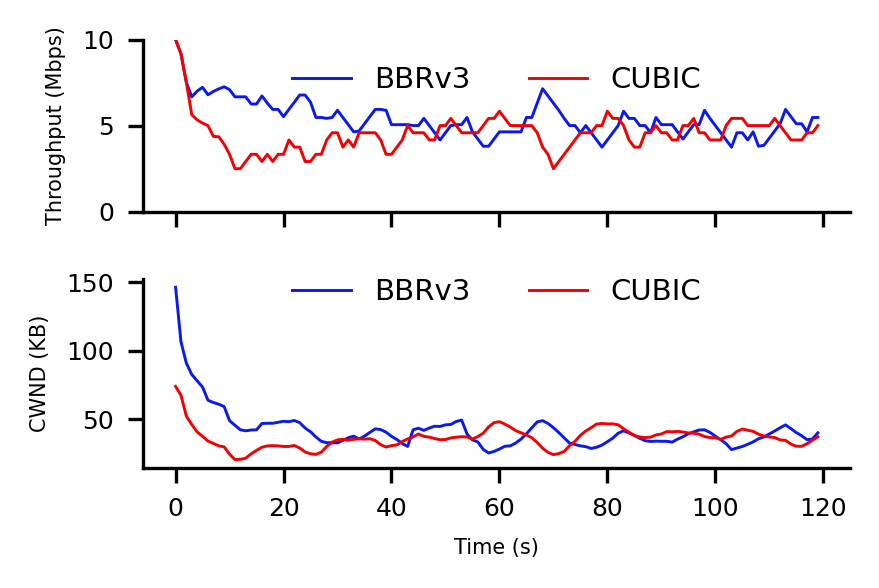}
        \caption{PFIFO}
        \label{fig:pfifo_thpt_dw}
    \end{subfigure}
    \hfill
    \begin{subfigure}[b]{0.15\textwidth}
        \centering
        \includegraphics[width=\textwidth]{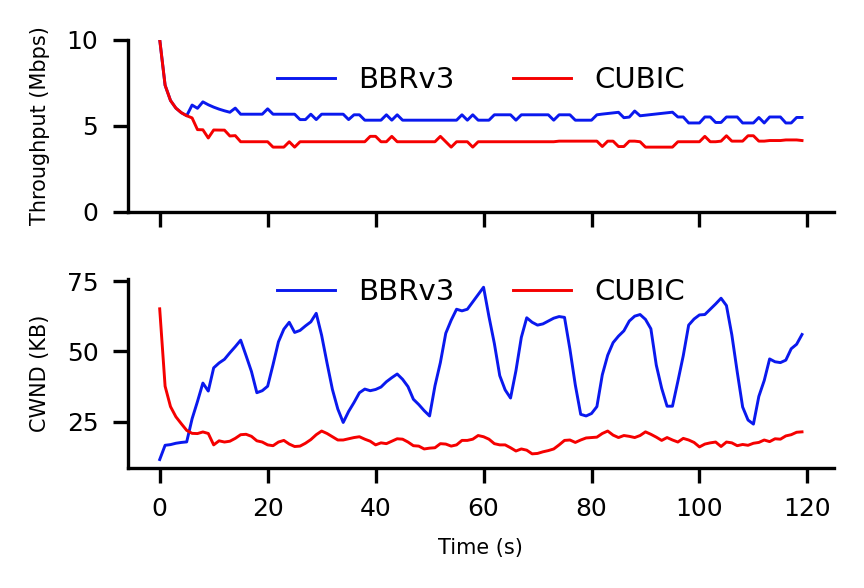}
        \caption{FQ-CoDel}
        \label{fig:fq-codel_thpt_dw}
    \end{subfigure}
    \hfill
    \begin{subfigure}[b]{0.15\textwidth}
        \centering
        \includegraphics[width=\textwidth]{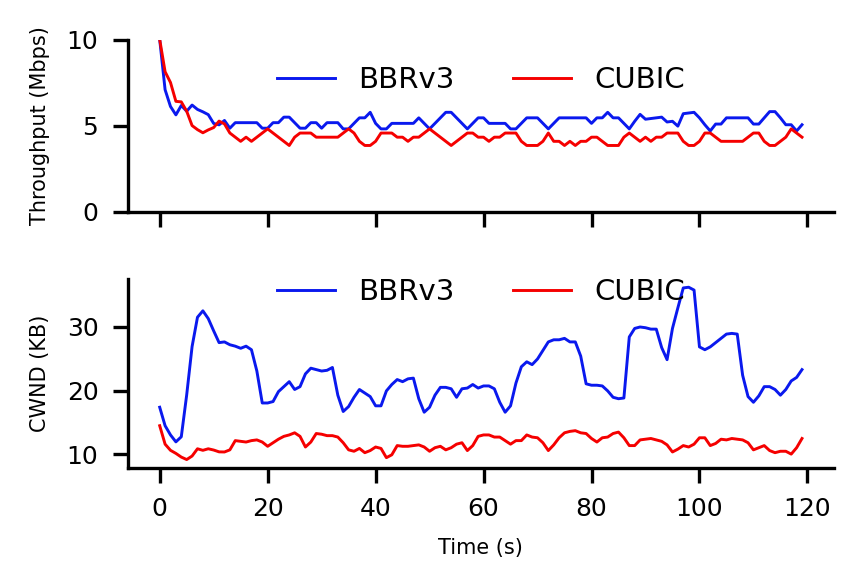}
        \caption{CAKE}
        \label{fig:cake_thpt_dw}
    \end{subfigure}
 
    \caption{Downlink throughput (top) and congestion window (bottom) dynamics for CUBIC (red) and BBRv3 (blue) under PFIFO, FQ-CoDel, and CAKE over a 10 Mbps Wi-Fi link.}
    \label{fig:allthroughput_dw}
\end{figure}

Under PFIFO (Figure~\ref{fig:pfifo_thpt_dw}), BBRv3 often achieves slightly higher throughput than CUBIC, a reversal from the uplink results. This is consistent with the model: with the AP as the single sender, $\Theta_i$ varies less, allowing BBRv3 to maintain a more accurate bandwidth estimate and execute its \textit{ProbeBW} cycle (Eq.~\ref{eq:w_pbw}) more effectively. Meanwhile, CUBIC’s loss-driven adjustments become less dominant, since PFIFO’s queue grows more gradually in downlink than in uplink. The $cwnd$ traces show BBRv3 opening its window more aggressively, reflecting improved estimation of bottleneck capacity. However, PFIFO still exhibits substantial delay inflation due to unregulated queue buildup, aligning with Eq.~\ref{eq:queue_dynamics} and the FIFO drop behavior in Eq.~\ref{eq:di_fifo}.

With FQ-CoDel (Figure~\ref{fig:fq-codel_thpt_dw}), throughput increases for both CCAs, but BBRv3 gains a larger advantage. FQ-CoDel isolates flows into virtual queues, ensuring each receives a stable service rate $\Theta_i$ while preventing excessive queuing delay (Eq.~\ref{eq:queue_dynamics}), consistent with CoDel’s sojourn-time rule in Eq.~\ref{eq:drop_fq}. This stabilizes RTT samples (Eq.~\ref{eq:total_rtt}), enabling BBRv3’s model-driven probing to track the true bottleneck rate accurately. As a result, BBRv3 consistently outperforms CUBIC and maintains a wider $cwnd$. Both flows operate close to the link’s efficient operating point, demonstrating how delay-regulating AQM optimally supports model-based CCAs.

Under CAKE (Figure~\ref{fig:cake_thpt_dw}), both flows stabilize
near 5\,Mbps with minimal variability, showing high fairness
and full utilization of the 10\,Mbps downlink. CAKE’s per-flow
fairness and shaping enforce a precise service schedule, keeping
queue lengths bounded and RTT variations minimal, closely
matching the steady-state regime predicted by the analytical
model. BBRv3 continues to probe more aggressively visible as
wider \texttt{cwnd} excursions but CAKE’s shaping prevents these
probes from disrupting coexistence, consistent with the 
per-host fairness allocation in Eq.~\ref{eq:alpha_cake} and the 
adaptive marking behavior in Eq.~\ref{eq:si_cake}. 
The resulting throughput is smooth and consistent across flows,
illustrating the suitability of CAKE’s structured queuing for
mixed-CCA downlink operation.

\begin{figure}[htbp]
    \centering
    \begin{subfigure}[b]{0.15\textwidth}
        \centering
        \includegraphics[width=\textwidth]{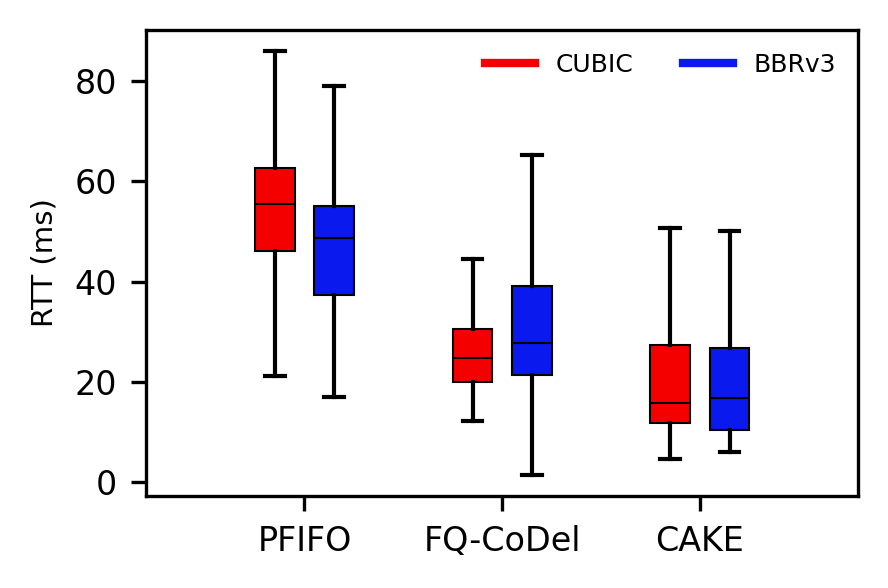}
        \caption{DL RTT}
        \label{fig:rtt_dw}
    \end{subfigure}
    \hfill
    \begin{subfigure}[b]{0.15\textwidth}
        \centering
        \includegraphics[width=\textwidth]{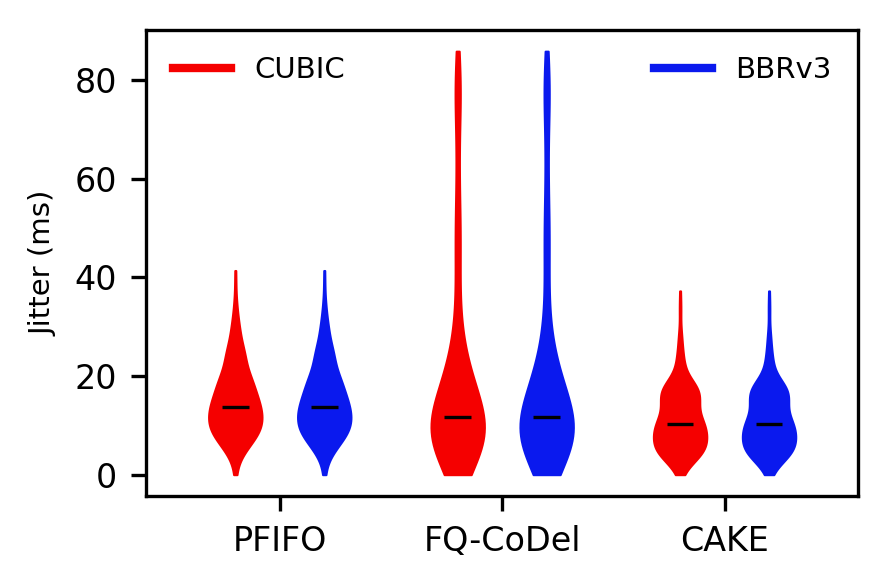}
        \caption{DL Jitter}
        \label{fig:jitter_dw}
    \end{subfigure}
    \hfill
    \begin{subfigure}[b]{0.17\textwidth}
        \centering
        \includegraphics[width=\textwidth]{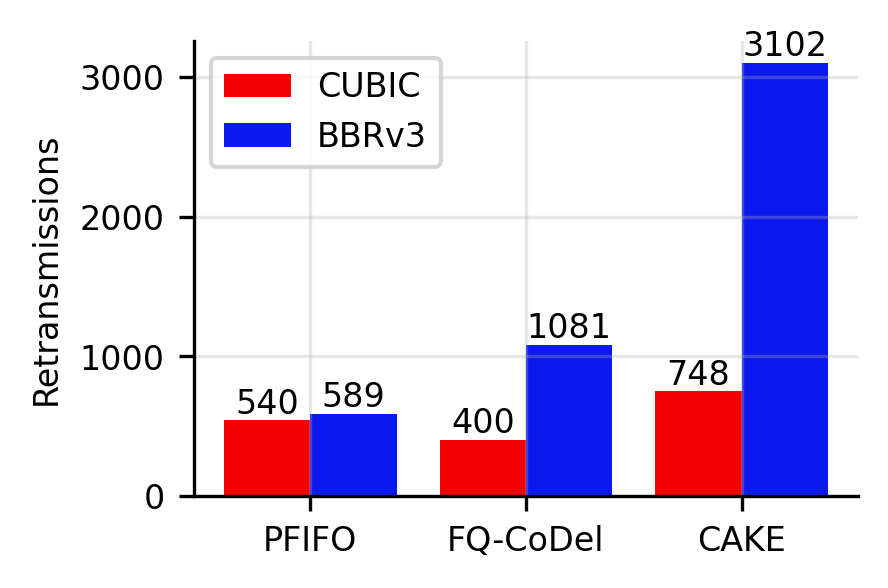}
        \caption{DL Retransmission}
        \label{fig:retransmission_dw}
    \end{subfigure}
 
    \caption{RTT, jitter, and retransmission statistics for downlink CUBIC and BBRv3 across AQMs.}
    \label{fig:3in1_dw}
\end{figure}

Figure~\ref{fig:rtt_dw} shows that PFIFO induces the highest RTTs ($\sim$50,ms) due to uncontrolled queue buildup, consistent with Eq.~\ref{eq:queue_dynamics}. FQ-CoDel reduces the median RTT by roughly half and significantly tightens its spread, while CAKE yields the lowest RTT ($<20$ms) with the most compact distribution. Jitter (Figure~\ref{fig:jitter_dw}) follows the same ordering, with CAKE providing the most stable delay profile. These results indicate that delay performance is governed primarily by the queue management discipline: the direction of traffic (uplink versus downlink) has only a marginal effect compared to the dominant influence of the AQM mechanism.

Figure~\ref{fig:retransmission_dw} reports retransmission behavior across the three AQMs. CUBIC generally incurs fewer retransmissions because its congestion window collapses sharply after a loss, reducing subsequent drops. BBRv3, however, triggers more frequent retransmissions especially under CAKE because its pacing continues probing even when CAKE’s shaping limits the effective service rate $\Theta_i$. This behavior is consistent with the queue dynamics in Eqs.~\ref{eq:si_cake}--\ref{eq:cake_dynamics}, where the adaptive drop/mark probability $m_i^{\mathrm{cake}}(t)$ governs the fraction of packets dropped to enforce fairness. Elevated $m_i^{\mathrm{cake}}(t)$ under aggressive probing increases retransmission events, but these losses are \textbf{not} a sign of instability; they reflect CAKE’s deliberate regulation of overly aggressive flows to maintain bounded queues and fair sharing.

\textit{\textbf{Takeaway:}  
Downlink flows exhibit more stable contention than uplink because the AP is the sole transmitter, but queue discipline still critically shapes fairness and delay. PFIFO suffers from queue inflation and inconsistent sharing, FQ-CoDel enhances fairness and supports BBRv3's model-driven probing, and CAKE provides the most controlled, low-latency, and balanced performance while restricting excessive probing through deliberate shaping.}

\subsection{Bidirectional Scenario}
\label{subsec:bidir_model_q1}

\begin{figure}[htbp]
\centering
\begin{subfigure}[b]{0.48\textwidth}
\centering
\includegraphics[width=\textwidth]{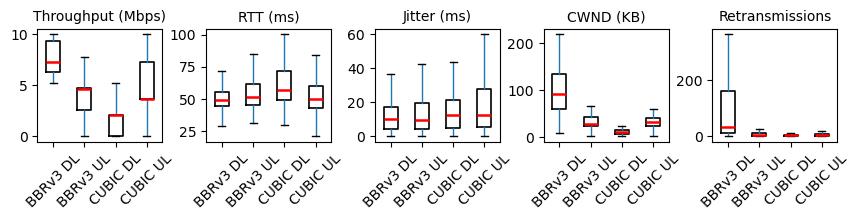}
\caption{BBRv3 vs.\ CUBIC in PFIFO}
\label{fig:fifo_bidr}
\end{subfigure}
\hfill
\begin{subfigure}[b]{0.48\textwidth}
\centering
\includegraphics[width=\textwidth]{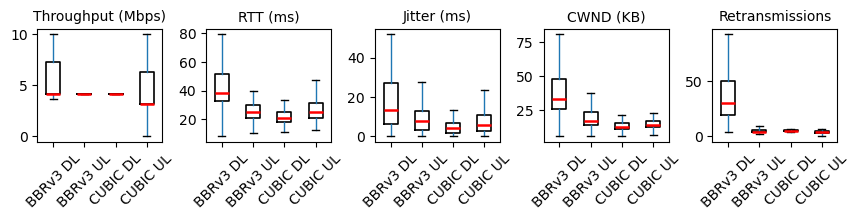}
\caption{BBRv3 vs.\ CUBIC in FQ-CoDel}
\label{fig:fqcodel_bidr}
\end{subfigure}
\hfill
\begin{subfigure}[b]{0.48\textwidth}
\centering
\includegraphics[width=\textwidth]{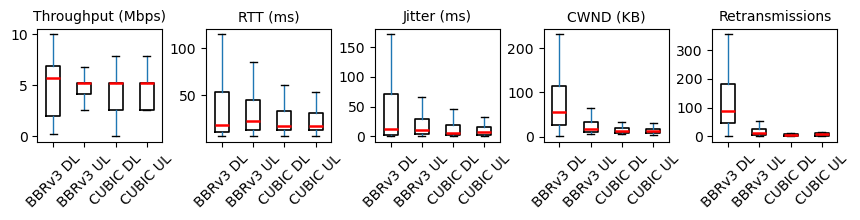}
\caption{BBRv3 vs.\ CUBIC in CAKE}
\label{fig:cake_bidr}
\end{subfigure}
\caption{Experimental bidirectional TCP performance over Wi-Fi under different AQMs: (a) PFIFO, (b) FQ-CoDel, and (c) CAKE.}
\label{fig:bidir_model_q1}
\end{figure}

We evaluate TCP performance under simultaneous upload and download flows to represent realistic bidirectional traffic. The results shown in Figure~\ref{fig:bidir_model_q1} are purely experimental measurements. However, the system model presented in Section~\ref{sec:system_model} provides a framework to interpret and explain these observations.

Experimental results (Figure~\ref{fig:fifo_bidr}) show that bidirectional traffic suffers from high RTT variability and erratic throughput. According to the system model (Section~\ref{sec:cross_layer}), these observations can be explained by queue dynamics in Eq.~\eqref{eq:queue_dynamics}: when the aggregate sending rate $x_i(t)$ of both upload and download flows exceeds the available MU--OFDMA service rate $\Theta_i$ (Eq.~\eqref{eq:per_STA-throughput}), queues $q_i(t)$ inflate rapidly, resulting in the observed bufferbloat and high RTT $\tau_i(t)$ (Eq.~\eqref{eq:total_rtt}). The BBRv3 \textit{ProbeBW} behavior (Eq.~\eqref{eq:w_pbw}--\eqref{eq:x_pbw}) further exaggerates this effect, as inflated RTT estimates temporarily allow over-inflated congestion windows, triggering retransmissions and erratic throughput.

Under FQ-CoDel (Figure~\ref{fig:fqcodel_bidr}), experimental throughput is more stable and RTT is significantly reduced. The model provides insight: per-flow service weights $\alpha_i^{\mathrm{fq}}(t)$ (Eq.~\eqref{eq:si_fq}) and CoDel drop/mark function $d_i^{\mathrm{fq}}(t)$ (Eq.~\eqref{eq:drop_fq}) prevent individual queues from growing beyond the target sojourn time. The plotted experimental data confirm that $q_i(t)$ is effectively regulated, maintaining the observed RTT near $\tau_i^{\min}$, in agreement with the analytical model.

\medskip

CAKE (Figure~\ref{fig:cake_bidr}) represents the state-of-the-art in queue management for bidirectional Wi-Fi traffic, delivering consistently high and balanced performance. Experimental results show throughput tightly clustered near the bottleneck capacity, RTTs approaching propagation delay with minimal jitter, and retransmissions near zero, reflecting near-perfect utilization and effective bufferbloat mitigation. These outcomes arise from CAKE’s fine-grained per-flow and per-host fairness, advanced shaping, and enhanced CoDel algorithm, which provide early congestion signaling often via ECN marks without hard packet drops. Congestion window values converge around the calculated bandwidth-delay product, indicating near-ideal in-flight data volumes with minimal oscillation. Per-host fairness $\phi_{\mathrm{host}}(h(i))$ and adaptive marking $m_i^{\mathrm{cake}}(t)$ (Eq.~\eqref{eq:alpha_cake}--\eqref{eq:cake_dynamics}) explain the observed low RTTs, minimal jitter, and high throughput, while the system model justifies why queues remain shallow and both uplink and downlink flows achieve near-optimal utilization.

\noindent\textit{\textbf{Takeaway:}} 
\textit{The experimental bidirectional TCP results are supported by the cross-layer analytical model: PFIFO saturates queues, FQ-CoDel restores fairness and reduces RTT, and CAKE achieves near-optimal utilization and minimal latency, consistent with the interactions predicted by Eqs.~\eqref{eq:queue_dynamics}--\eqref{eq:cake_dynamics}.}

\subsection{Practical Implications and Recommendations} \label{implication}

Our results have several practical implications for deploying BBRv3 in Wi-Fi networks:

\begin{itemize}
    \item While BBRv3 achieves improved fairness and responsiveness compared to earlier versions~\ref{bbr_advancement}, it can trigger high retransmission rates under CAKE~\ref{fig:retransmission_up} and \ref{fig:retransmission_dw} due to aggressive bandwidth probing. In latency-sensitive applications (e.g., VoIP or streaming), these retransmissions could degrade user experience.
    
    \item Network operators deploying BBRv3 should pair it with modern AQMs such as CAKE or FQ-CoDel to prevent severe unfairness and bufferbloat, especially in mixed-CCA environments~\ref{fig:allthroughput_up}, \ref{fig:allthroughput_dw} and \ref{fig:bidir_model_q1}.
    
    \item Tuning parameters such as \texttt{pacing\_gain} and \texttt{cwnd\_gain} for Wi-Fi scenarios could help mitigate oscillatory behavior and reduce retransmissions, suggesting a need for further algorithm refinements.
    
    \item For residential users, sticking with traditional loss-based CCAs such as CUBIC might still be advisable in unmanaged Wi-Fi environments~\ref{bbr_advancement} lacking proper AQM support.
\end{itemize}

These insights highlight that queue-aware designs remain essential for realizing the full benefits of model-based CCAs such as BBRv3 in wireless networks.

\section{Conclusions and Future Work} \label{conclusion}

In this paper, we present the first comprehensive empirical and analytical study of TCP BBRv3 over Wi-Fi~6 in modern AQM-enabled environments. Using a real-world Wi-Fi 6 testbed with AQM-enabled commodity home gateways, we showed that BBRv3 lowers standing queues and improves fairness compared to earlier BBR versions. However, its performance remains sensitive to wireless rate variability and AQM behavior. We also discovered a new retransmission issue caused by pacing and delivery misalignment during \textit{ProbeBW}, especially under CAKE, which reduces throughput even when losses are low. This behavior does not appear in wired networks.

To explain these findings, we developed a cross-layer model that links MU-OFDMA scheduling, queue evolution, and BBRv3’s pacing and inflight control. The model shows how changes in service rates and queue delay can distort bandwidth estimates and hinder \textit{ProbeBW} transitions. Overall, BBRv3 improves latency and fairness, but its current design assumptions are not always suitable for highly variable wireless links.

Future work will focus on four directions. First, we plan to adapt BBRv3’s pacing and in-flight logic to account for MU-OFDMA variability and changing delivery rates. Second, AQM and model-based CCAs should be designed together to balance latency control and stable rate estimation. Third, exposing MAC-layer information such as RU allocation and contention level may help to improve bandwidth estimation. Finally, we will evaluate these ideas on Wi-Fi 7, multi-link systems, and diverse client devices to test whether the findings hold and generalize to next-generation Wi-Fi networks.

\bibliographystyle{IEEEtran}
\bibliography{refs}

\end{document}